\documentclass[journal]{IEEEtran}
\usepackage{graphicx}
\usepackage{amsmath,amssymb,mathtools}
\usepackage{caption}
\usepackage{algorithm}
\usepackage{algpseudocode}
\usepackage{subcaption}
\usepackage{tabularx}
\usepackage{array}
\usepackage{dblfloatfix}
\usepackage{orcidlink}

\usepackage{cite}
\usepackage{booktabs}
\usepackage{siunitx}
\usepackage{microtype}
\usepackage{float}
\allowdisplaybreaks
\title{Online Model Predictive Control for Trajectory and Beamforming Optimization in UAV-Enabled URLLC}
\author{Asim Ihsan \,\orcidlink{0000-0001-7491-7178},
Muhammad Asif \,\orcidlink{0000-0002-9699-1675},
Ali Arshad Nasir\,\orcidlink{0000-0001-5012-1562}, Senior Member, IEEE,
Khaled M. Rabie\,\orcidlink{0000-0002-9784-3703}, Senior Member, IEEE,
and Wali Ullah Khan\,\orcidlink{0000-0003-1485-5141}
\thanks{Corresponding author: Khaled M. Rabie}
\thanks{Asim Ihsan is with the Interdisciplinary Research Center for Communication Systems and Sensing (IRC-CSS), King Fahd University of Petroleum and Minerals (KFUPM), Dhahran 31261, Saudi Arabia (Email: asim.ihsan@kfupm.edu.sa).}
\thanks{Muhammad Asif is with the School of Electrical Engineering, Tongling University, Tongling 244061, China (Email: masif@tlu.edu.cn).}
\thanks{Ali Arshad Nasir is with the Department of Electrical Engineering and the Center for Communication Systems and Sensing, King Fahd University of Petroleum and Minerals (KFUPM), Dhahran 31261, Saudi Arabia (Email: anasir@kfupm.edu.sa).}
\thanks{Khaled M. Rabie is with Department of Computer Engineering and Research Center for Communication Systems and Sensing, King Fahd University of Petroleum and Minerals (KFUPM), Dhahran 31261, Saudi Arabia (Email: k.rabie@kfupm.edu.sa).}
\thanks{Wali Ullah Khan is with the Interdisciplinary Centre for Security, Reliability, and Trust, University of Luxembourg, 1855 Luxembourg City, Luxembourg (Email: waliullah.khan@uni.lu).}
}

\begin{document}

\maketitle

\begin{abstract}
This paper investigates joint trajectory and active beamforming design for unmanned aerial vehicle (UAV)-enabled ultra-reliable low-latency communication (URLLC) systems under finite blocklength (FBL) transmission. Unlike conventional Shannon-capacity formulations, the FBL regime introduces a signal-to-interference-plus-noise ratio (SINR)-dependent dispersion penalty that significantly increases the sensitivity of reliability to mobility-induced channel variations. To address this challenge, we develop a propulsion-aware model predictive control (MPC) framework that performs receding-horizon joint trajectory and multi-user beamforming optimization while explicitly enforcing FBL-based rate constraints. The resulting long-horizon non-convex problem is decomposed into beamforming and trajectory subproblems using an alternating optimization framework, where concave surrogate is constructed for the Shannon capacity term and convex approximations are derived for the dispersion component and the nonlinear propulsion power model to obtain tractable convex subproblems solved iteratively. Compared with an offline MPC scheme, where the predictive control problem is solved once over the entire mission horizon without receding-horizon feedback, and a conventional offline optimization approach that jointly optimizes trajectory and beamforming without an MPC structure, the proposed closed-loop framework achieves disturbance-resilient mission completion in the presence of UAV position disturbances. Simulation results further demonstrate that, compared with conventional beamforming strategies such as maximum ratio transmission (MRT) and equal-power allocation, the proposed interference-aware design significantly enhances URLLC reliability under stringent minimum rate requirements. In addition, the results quantify the impact of antenna scaling, transmit power, and transmission time on FBL performance, providing practical insights for reliability-centric UAV-enabled wireless networks in 5G and beyond.
\end{abstract}


\section{INTRODUCTION}

Unmanned aerial vehicles (UAVs) equipped with wireless transceivers can operate as airborne base stations (UAV-BSs) to dynamically enhance coverage and throughput in scenarios where terrestrial infrastructure is congested, insufficient, or unavailable \cite{9220821}. UAV-BSs have emerged as rapid and flexible solutions for disaster relief and emergency response, where ground BSs may be damaged and reliable connectivity is required to support mission-critical communications \cite{9541037}. By exploiting their three-dimensional mobility and favorable air-to-ground propagation characteristics, UAV-BSs can be deployed on demand to restore connectivity, mitigate coverage gaps, and support public safety operations \cite{10381632}. Consequently, UAV-BS-enabled systems represent a promising complement to terrestrial networks, particularly in scenarios requiring rapid reconfiguration and high reliability.

The communication services demanded in mission-critical UAV-BS deployments inherently impose extremely stringent latency and reliability requirements, which fall under the category of ultra-reliable low-latency communication (URLLC) in fifth-generation (5G) and beyond wireless networks \cite{10680585,10711854}. URLLC applications, including emergency response, remote control, and industrial automation, typically require end-to-end latency of the order of milliseconds and packet error probabilities as low as $10^{-5}$ or below \cite{9246274}. These requirements fundamentally distinguish URLLC from conventional enhanced mobile broadband (eMBB) services and pose significant challenges for wireless system design, particularly in UAV-BS-enabled networks where channel conditions vary rapidly due to mobility and environmental dynamics. To meet ultra-low latency constraints, URLLC systems rely on short-packet transmissions, which invalidate the classical infinite blocklength assumption underlying the Shannon capacity theory \cite{9080500}. In the resulting finite blocklength (FBL) regime, decoding errors can no longer be neglected, and the achievable transmission rate is governed by a nontrivial tradeoff among rate, latency, and reliability \cite{5452208}. Consequently, accurate performance modeling and resource optimization for URLLC systems must explicitly incorporate FBL information-theoretic models rather than conventional capacity-based formulations \cite{11214543}. In particular, under FBL transmission, the achievable rate becomes an explicit function of the instantaneous signal-to-interference-plus-noise ratio (SINR), while latency and reliability are reflected through blocklength and decoding error probability, respectively. Since instantaneous SINR depends on the position of the UAV and the channel realization, mobility-induced channel variations directly affect the feasibility of URLLC constraints in UAV-enabled systems. Unlike asymptotic capacity formulations, the dispersion term in the FBL rate expression amplifies the impact of SINR fluctuations, rendering reliability constraints highly sensitive to mobility-induced dynamics. As a result, trajectories that are feasible under Shannon-rate assumptions may become infeasible under FBL constraints, fundamentally altering the spatial positioning strategy of the UAV.

Beyond these fundamental considerations, previous studies have demonstrated that exploiting UAV mobility through trajectory optimization can significantly improve communication performance compared to static aerial UAV deployments and conventional terrestrial base stations, due to flexible spatial positioning and improved channel conditions \cite{8247211,7572068}. By dynamically adjusting UAV trajectories, reliable communication links can be maintained for spatially distributed ground users, making UAV-assisted systems particularly suitable for applications that demand low latency and high reliability \cite{9682551,10564098}. As a result, joint UAV trajectory optimization and physical-layer transmission design have emerged as a key approach to improve the reliability and efficiency of UAV-assisted wireless communication systems, especially under URLLC requirements \cite{SHENG2023102063, hu2025energy}.
\begin{table*}[!t]
\centering
\caption{Comparison of Joint Trajectory and Transmission Optimization Works for UAV-Assisted URLLC Systems}
\label{tab:comparison}
\renewcommand{\arraystretch}{1.2}
\small
\setlength{\tabcolsep}{4pt}

\begin{tabular}{lcccccccc}
\hline
\textbf{Work} 
& \textbf{Trajectory} 
& \textbf{Active BF} 
& \textbf{URLLC} 
& \textbf{FBL} 
& \textbf{Dispersion} 
& \textbf{MPC} 
& \textbf{Propulsion} 
& \textbf{Per-step} \\
\hline

Zhang \textit{et al.} \cite{9834715}
& \checkmark 
& $\times$ (Passive IRS) 
& \checkmark 
& \checkmark 
& $\times$ 
& $\times$ 
& $\times$ 
& $\times$ \\

Zhang \textit{et al.} \cite{9912224}
& \checkmark 
& \checkmark 
& \checkmark 
& \checkmark 
& $\times$ 
& $\times$ 
& $\times$ 
& $\times$ \\

Qin \textit{et al.} \cite{10680585}
& \checkmark 
& \checkmark 
& \checkmark 
& $\times$ 
& $\times$ 
& $\times$ 
& $\times$ 
& $\times$ \\

Sheng \textit{et al.}  \cite{SHENG2023102063}
& \checkmark 
& $\times$ 
& \checkmark 
& \checkmark 
& $\times$ 
& $\times$ 
& $\times$ 
& $\times$ \\

Hu \textit{et al.} \cite{hu2025energy}
& \checkmark 
& $\times$ (Passive IRS) 
& \checkmark 
& \checkmark 
& $\times$ (Dispersion $\approx 1$) 
& $\times$ 
& \checkmark 
& $\times$ \\

Chen \textit{et al.}  \cite{10044099}
& \checkmark 
& $\times$ 
& \checkmark 
& \checkmark 
& $\times$ 
& $\times$ 
& \checkmark 
& $\times$ \\

\textbf{This Work} 
& \checkmark 
& \checkmark 
& \checkmark 
& \checkmark 
& \checkmark 
& \checkmark 
& \checkmark 
& \checkmark \\

\hline
\multicolumn{9}{p{\textwidth}}{\footnotesize
\textit{Note:} 
Per-step: Per-time-step URLLC feasibility.
}
\end{tabular}
\end{table*}

Recent works have investigated joint trajectory and transmission optimization for UAV-assisted URLLC systems under different reliability modeling frameworks. Zhang \textit{et al.} \cite{9834715} and Zhang \textit{et al.} \cite{9912224} consider joint UAV trajectory and transmission design under FBL modeling with statistical delay and error-rate bounded QoS constraints, incorporating passive IRS beamforming or MIMO-based strategies. Sheng \textit{et al.} \cite{SHENG2023102063} and Chen \textit{et al.} \cite{10044099} integrate short-packet FBL transmission into UAV-assisted secure communication and data collection frameworks, while Hu \textit{et al.} \cite{hu2025energy} incorporate FBL modeling and propulsion energy considerations in IRS-assisted UAV systems. Qin \textit{et al.} \cite{10680585} develop URLLC-aware trajectory planning and active beamforming strategies using queue-based reliability adaptation. Although several of these works adopt FBL rate expressions, the SINR-coupled dispersion term is often approximated or handled implicitly, without explicitly deriving tractable convex surrogate reformulations within an iterative optimization framework. Moreover, most existing approaches formulate the joint trajectory and transmission optimization as offline full-horizon problems executed in an open-loop manner, without incorporating receding-horizon state feedback to correct deviations between the planned and actual UAV trajectory caused by motion disturbances and actuation errors. As summarized in Table~\ref{tab:comparison}, existing studies rarely address URLLC under FBL within an MPC framework for joint UAV beamforming and trajectory optimization with propulsion energy modeling and mission constraints from a predefined start to a destination. To the best of our knowledge, no prior work integrates these elements for UAV-assisted URLLC within a unified MPC framework.

In mission-oriented UAV-BS scenarios, the UAV must travel from a prescribed starting location to a designated destination while continuously providing communication services \cite{7888557,9453799}. Such goal-oriented trajectory planning introduces a long-horizon optimization problem in which trajectory decisions affect both mission completion efficiency and communication performance. Under URLLC constraints based on FBL, mobility-induced channel variations directly influence the instantaneous SINR and thus the achievable FBL rate at each discrete time step. Ensuring per-time-step reliability feasibility therefore necessitates adaptive coordination between trajectory evolution and transmission design.

To enable such adaptive and constraint-aware coordination, model predictive control (MPC) provides a natural framework. MPC is a well-established constrained optimal control methodology that repeatedly solves a finite-horizon optimization problem and implements the first control action in a receding-horizon manner, thereby incorporating real-time state feedback and disturbance rejection \cite{MAYNE2000789,mayne2009model}. In the context of UAV systems, MPC has been widely used for constraint-aware navigation, trajectory tracking, and fault-tolerant flight control, where state and input constraints must be explicitly satisfied under model uncertainties and external disturbances \cite{9145644, 9834096, nan2022nonlinear }. Beyond pure flight control, MPC has also been adopted in UAV-enabled communication and energy-aware networking systems to jointly optimize mobility and communication objectives under dynamic environmental conditions \cite{9453799, 11069265}. These prior works demonstrate the effectiveness of receding-horizon control in managing coupled motion and resource constraints. However, existing MPC-based UAV communication designs primarily rely on Shannon-capacity formulations and do not explicitly incorporate FBL dispersion penalties governing per-time-step URLLC feasibility. This motivates the development of a propulsion-aware MPC framework that integrates SINR-coupled FBL rate constraints within a closed-loop predictive optimization structure. This feedback-driven structure mitigates performance degradation caused by trajectory deviations and model mismatches that cannot be corrected under single-shot open-loop optimization.

The main contributions of this paper are summarized as follows:
\begin{itemize}
    \item First, we establish a closed-loop, propulsion-aware predictive optimization framework that embeds FBL rate constraints directly into joint UAV trajectory and multi-user beamforming control. By integrating state-feedback-driven receding-horizon optimization, the proposed design enforces strict per-time-step URLLC feasibility under mobility-induced SINR variations, enabling reliability-centric aerial communication under dynamic and disturbance-prone operation.
    \item Moreover, we explicitly model the coupling between UAV mobility, instantaneous SINR, and FBL dispersion, revealing the heightened reliability sensitivity to mobility-induced channel fluctuations in the short-packet regime. By incorporating this SINR-dependent dispersion structure into the MPC optimization process, the framework enables predictive, dispersion-aware trajectory–beamforming adaptation under FBL transmission.
    \item Additionally, we incorporate a realistic nonlinear propulsion power model into the proposed closed-loop MPC framework with explicit per-time-step FBL rate constraints. The resulting formulation jointly optimizes the UAV trajectory, velocity, and beamforming while accounting for propulsion energy consumption, capturing the coupling among mobility-induced SINR dynamics, dispersion-sensitive reliability, and propulsion power.
    \item Furthermore, through receding-horizon state feedback, the framework corrects UAV position deviations occurring during flight execution, preventing violations of FBL reliability constraints and enabling robust URLLC mission performance beyond open-loop designs.
    \item Finally, we develop a tractable convex reformulation of the SINR-coupled FBL rate constraints and nonlinear propulsion model within the joint beamforming–trajectory optimization framework. By constructing tight surrogate bounds for both the Shannon term and the dispersion penalty from the beamforming and UAV mobility perspectives, the proposed design enables efficient iterative optimization while preserving feasibility under mobility-dependent channel variations.
    
\end{itemize}

\section{SYSTEM MODEL}

\begin{figure}[ht!]
    \centering
    \includegraphics[width=3.5 in]{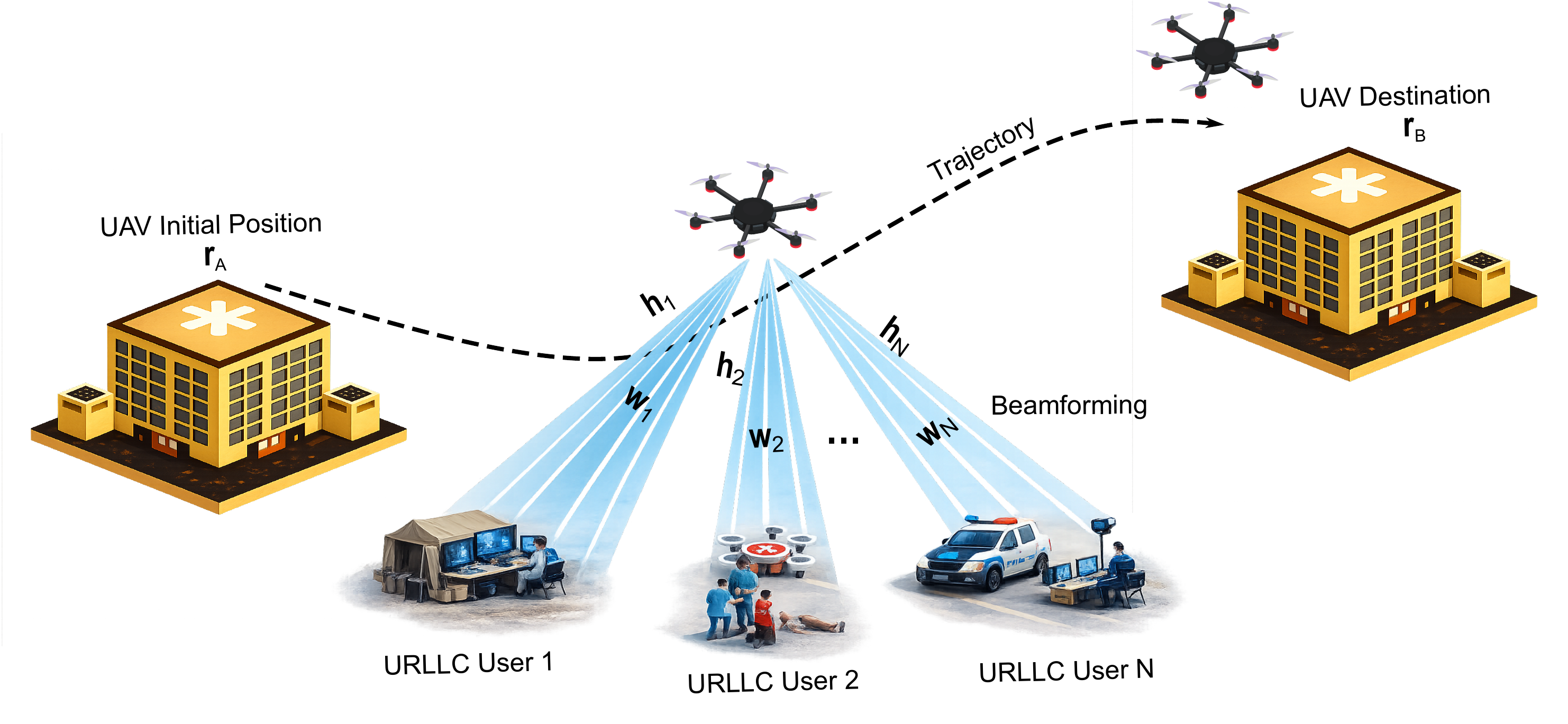}
    \caption{Illustration of System Model.}
    \label{fig:system_model}
    \label{System Model}
\end{figure}

We consider a downlink communication system in which a UAV equipped with a uniform linear array (ULA) of $M$ antennas serves $N$ single-antenna users randomly distributed in a three-dimensional (3D) coverage area. Each user $n$ is located at $\mathbf{u}_n = (x_n, y_n, z_n)$. The horizontal coordinates $(x_n, y_n)$ are uniformly distributed within a corridor-shaped region aligned with the UAV mission path from $\mathbf{r}_A$ to $\mathbf{r}_B$, with corridor width $W_c$. The vertical coordinate $z_n$ represents the user height and is uniformly distributed within the interval $[0.5, 3]$ meters to model realistic device elevations. The overall UAV-enabled URLLC system and mission-oriented trajectory configuration are illustrated in Fig.~\ref{System Model}.

The UAV flies from a designated starting point $\mathbf{r}_A = (x_A, y_A, z_A)^T$ to a specified endpoint $\mathbf{r}_B = (x_B, y_B, z_B)^T$, with its trajectory discretized into $T$ time steps. The UAV position and velocity at time step $t$ are denoted by $\mathbf{r}(t) = (r_x(t), r_y(t), r_z(t))$ and $\mathbf{v}(t) = (v_x(t), v_y(t), v_z(t))$, respectively. The UAV motion is modeled in discrete time:
\begin{equation}
\mathbf{r}(t+1) = \mathbf{r}(t) + \mathbf{v}(t) t_c,
\label{eq:1}
\end{equation}
where $t_c$ is the duration of each time step. The time-step length is selected according to UAV mobility dynamics. In particular, with maximum horizontal speed $V_{\max}$, the UAV displacement within one time step is upper bounded by $V_{\max} t_c$, ensuring sufficiently fine-grained trajectory discretization relative to typical communication link distances \cite{10680585,8247211}. Although URLLC packets typically have millisecond-level transmission durations \cite{9246274}, multiple short packets may be transmitted within each UAV time step. Since the multi-path effects due to small-scale fading can be averaged out within each time step \cite{10680585, 9034493}, the channel evolution across time steps is primarily governed by the geometry-dependent large-scale components.

The UAV dynamics are subject to practical constraints:
\begin{align}
&Z_{\min} \le r_z(t) \le Z_{\max}, \label{eq:2}  \\
&\| (v_x(t), v_y(t)) \| \le V_{\max}, \label{eq:3}\\
&v_z(t) \le U_{\max}, \label{eq:4}\\
&\| \mathbf{v}(t+1) - \mathbf{v}(t) \| \le a_{\max} t_c, \label{eq:5}
\end{align}
where $Z_{\min}$ and $Z_{\max}$ denote the minimum and maximum flight altitudes, $V_{\max}$ and $U_{\max}$ are the maximum horizontal and vertical speeds, and $a_{\max}$ is the maximum UAV acceleration. These constraints ensure the UAV's motion remains physically feasible.

During flight, the UAV's total power consumption consists of propulsion and communication-related components. The propulsion power at time step $t$ is modeled as \cite{9453799}:
\begin{equation}
\begin{aligned}
P_{\mathrm{prop}}(\mathbf{v}(t)) &= \frac{W^2}{\sqrt{2} \rho S} \Big( \|\mathbf{v}_h(t)\|^2 + \sqrt{\|\mathbf{v}_h(t)\|^4 + 4 V_h^4} \Big)^{-1/2} \\
&\quad + W \max\big(v_z(t), 0\big) + \frac{\zeta \rho S}{8} \|\mathbf{v}_h(t)\|^3, \label{eq:6}
\end{aligned}
\end{equation}
where $\mathbf{v}_h(t) = [v_x(t), v_y(t)]^T$ is the horizontal velocity and $v_z(t)$ is the vertical velocity. 
Here, $W$ denotes the UAV weight, $\rho$ is the air density, $S$ is the rotor disk area, $V_h = \sqrt{W / (2 \rho S)}$ is the hover parameter, and $\zeta$ is the profile drag coefficient. The term $W\max(v_z(t),0)$ accounts for additional power during climbing, while descent ($v_z(t)<0$) does not yield negative power, reflecting the practical assumption that rotary-wing UAVs do not perform regenerative energy recovery and require rotor thrust to maintain controlled flight. 

The communication-related power at time step $t$ is expressed as
\begin{equation}
P_{\mathrm{com}}(t) = \frac{1}{\eta} \sum_{n=1}^{N} \|\mathbf{w}_n(t)\|^2, \label{eq:7}
\end{equation}
where $\eta$ denotes the power amplifier efficiency, and $\mathbf{w}_n(t)$ is the beamforming vector for user $n$.  

The total UAV power consumption is then
\begin{equation}
P_{\mathrm{tot}}(t) = P_{\mathrm{prop}}(\mathbf{v}(t)) + P_{\mathrm{com}}(t). \label{eq:8}
\end{equation}
The total power consumption in Eq.~\eqref{eq:8} characterizes the instantaneous
power consumption of the UAV at each time step, accounting for both propulsion
and communication-related components. To evaluate the overall
energy usage over the mission duration, we define the UAV energy consumption
as the time integral of its total power consumption \cite{8663615}.
Under time step based UAV operation with a fixed sampling interval $t_c$, this
continuous-time formulation is approximated using a standard discretization.
Specifically, following prior UAV trajectory optimization works that compute
energy as the sum of per-segment power consumption multiplied by the
corresponding duration \cite{9775598}, the total energy consumption in
this work is approximated as
\begin{equation}
E \;\approx\; \sum_{t=1}^{T} P_{\mathrm{tot}}(t)\, t_c .
\label{eq:9}
\end{equation}

The large-scale fading between the UAV and user $n$ is characterized by a distance-dependent path loss:
\begin{equation}
\beta_n(t) = \frac{B_0}{d_n(t)^\rho}, \label{eq:10}
\end{equation}
where $B_0$ is the reference channel power at one meter and $\rho$ is the path loss exponent. Here, $d_n(t)$ denotes the three-dimensional Euclidean distance between the UAV and user $n$ at time step $t$, given by
\begin{align}
d_n(t) &= \|\mathbf{u}_n - \mathbf{r}(t)\|_2  \nonumber\\
       &= \sqrt{(x_n - r_x(t))^2 + (y_n - r_y(t))^2 + (z_n - r_z(t))^2}. \label{eq:11}
\end{align}
 The small-scale fading is modeled as Rician fading to capture both line-of-sight (LoS) and scattered non-line-of-sight (NLoS) components, resulting in the channel vector
\begin{equation}
\mathbf{h}_n(t) = \sqrt{\beta_n(t)} \Bigg( \sqrt{\frac{K_n}{K_n + 1}} \mathbf{a}(\theta_n(t)) + \sqrt{\frac{1}{K_n + 1}} \mathbf{g}_n(t) \Bigg), \label{eq:12}
\end{equation}
where $K_n$ is the Rician $K$-factor, $\mathbf{g}_n(t) \sim \mathcal{CN}(0, \mathbf{I}_M)$ represents the NLoS component, and $\mathbf{a}(\theta_n(t))$ is the steering vector of the UAV’s ULA corresponding to the angle of departure (AoD) $\theta_n(t)$. Assuming half-wavelength antenna spacing, the steering vector is
\begin{equation}
\mathbf{a}(\theta_n(t)) = [1, e^{j \pi \cos(\theta_n(t))}, \dots, e^{j (M-1) \pi \cos(\theta_n(t))}]^T, \label{eq:13}
\end{equation}
with
\begin{equation}
\cos(\theta_n(t)) = \frac{x_n - r_x(t)}{d_n(t)}. \label{eq:14}
\end{equation}

The UAV employs beamforming vectors $\mathbf{w}_n(t)$ to transmit symbols to each user. The received signal at user $n$ is
\begin{equation}
y_n(t) = \mathbf{h}_n^H(t) \mathbf{w}_n(t) s_n(t) + \sum_{k \ne n} \mathbf{h}_n^H(t) \mathbf{w}_k(t) s_k(t) + n_n(t), \label{eq:15}
\end{equation}
where $s_n(t)$ is the transmitted symbol for user $n$ with $\mathbb{E}[|s_n(t)|^2]=1$, and $n_n(t) \sim \mathcal{CN}(0,\sigma^2)$ denotes additive white Gaussian noise. The corresponding SINR is
\begin{equation}
\gamma_n(t) =\frac{S_n(t)}{I_n(t)}= \frac{|\mathbf{h}_n^H(t)\mathbf{w}_n(t)|^2}{\sum_{k \ne n} |\mathbf{h}_n^H(t) \mathbf{w}_k(t)|^2 + \sigma^2}. \label{eq:16}
\end{equation}

To account for URLLC, we adopt a FBL model. The achievable rate for user $n$ at time step $t$ is approximated as
\begin{equation}
R_{n}(t) =
\underbrace{\log\big(1+\gamma_{n}(t)\big)}_{ C_{n}(t)}
-
\underbrace{Q^{-1}(\varepsilon_{n}) \sqrt{\frac{V_{n}(t)}{L}}}_{ D_{n}(t)}, \label{eq:17}
\end{equation}

where $L$ is the blocklength, $\varepsilon_n$ is the target decoding error probability, $Q^{-1}(\cdot)$ denotes the inverse Gaussian Q-function, and the channel dispersion is
\begin{equation}
V_n(t) = \Big(1 - \frac{1}{(1 + \gamma_n(t))^2}\Big). \label{eq:18}
\end{equation}
Here, $C_n(t)$ denotes the Shannon capacity term, while $D_n(t)$ represents the FBL penalty.

This system model establishes the relationship between the UAV’s trajectory, the user distribution, the wireless channel characteristics, and the achievable rates, forming the foundation for the subsequent trajectory and resource optimization problem.

\section{PROBLEM FORMULATION AND ITS SOLUTION}
Based on the system model, our goal is to jointly optimize the UAV trajectory and the transmit beamforming in order to maximize the network performance, while satisfying practical motion and power constraints. 
Specifically, let $\mathbf r(t)$ and $\mathbf v(t)$ denote the UAV position and velocity at time step $t$, and let $\mathbf w_n(t)$ represent the beamforming vector for user $n$. The corresponding joint long-horizon optimization problem can then be formulated as
\begin{align}
\textbf{(P1)}\;
\max_{\{\mathbf r(t),\,\mathbf v(t),\,\mathbf w_n(t)\}_{t=1}^{T}}
& \sum_{t=1}^{T} \Bigg[
\psi_1\sum_{n=1}^{N} R_n(t)
- \psi_2 \|\mathbf r(t)-\mathbf r_B\|_2^2  \nonumber\\
&\quad - \psi_3 P_{\mathrm{prop}}(\mathbf{v}(t))
\Bigg] \nonumber\\
\text{s.t.}\quad
\text{C1:} &\ \mathbf r(t+1) = \mathbf r(t) + \mathbf v(t) t_c, \nonumber\\
\text{C2a:} &\ X_{\min} \le r_x(t) \le X_{\max}, \nonumber\\
\text{C2b:} &\ Y_{\min} \le r_y(t) \le Y_{\max}, \nonumber\\
\text{C2c:} &\ Z_{\min} \le r_z(t) \le Z_{\max}, \nonumber\\
\text{C3:} &\ \|\mathbf v_h(t)\| \le V_{\max}, \nonumber\\
\text{C4:} &\ |v_z(t)| \le U_{\max}, \nonumber\\
\text{C5:} &\ \|\mathbf v(t+1) - \mathbf v(t)\|
\le a_{\max} t_c, \nonumber\\
\text{C6:} &\ P_{\text{tot}}(t) \le P_{\max}, \nonumber\\
\text{C7:} &\ \frac{1}{\eta}\sum_{n=1}^{N}
\|\mathbf w_n(t)\|_2^2 \le P^{com}_{\max}, \nonumber\\
\text{C8:} &\ R_n(t) \ge R_n^{\min}, \quad \forall n.
\label{eq:19}
\end{align}
In (\textbf{P1}), the objective function consists of three parts. The first term, $\sum_{n=1}^N R_n(t)$, maximizes the aggregate FBL rates of all users, thereby enhancing the overall URLLC communication performance. The second term, $-\psi\|\mathbf{r}(t)-\mathbf{r}_B\|^2$, penalizes deviations from the target $\mathbf{r}_B$ and thus aims to optimize the UAV’s flight path from the starting point to the destination at the highest possible speed. The third term, $-\psi_3 P_{\text{prop}}(v(t))$,  penalizes the propulsion power consumption of the UAV, encouraging trajectory and velocity profiles that reduce propulsion energy expenditure. The constraints C1–C8 guarantee physical feasibility and communication compliance. Constraint C1 specifies the discrete-time kinematic model of the UAV. Constraint C2 restricts the UAV altitude within the feasible range $[Z_{\min},Z_{\max}]$. The velocity limits are captured in C3 and C4, which bound the horizontal and vertical speeds by $V_{\max}$ and $U_{\max}$, respectively. Constraint C5 enforces the maximum allowable acceleration to ensure smooth trajectory updates. Constraint C6 imposes the per-step power budget, which accounts for both propulsion power $P_{\text{prop}}(v(t))$ and communication power. Constraint C7 limits the communication transmit power through the beamforming vectors. Finally, constraint C8 enforces the per-user per-time-step URLLC quality-of-service requirement by guaranteeing $R_n(t) \ge R_n^{\min}$ for all users. Altogether, (\textbf{P1}) balances throughput maximization, mission completion, and propulsion energy consumption, while satisfying UAV mobility constraints and communication quality-of-service requirements. However, solving (\textbf{P1}) directly is not feasible in real time because it requires complete knowledge of future system states and leads to a large-scale non-convex problem. 

To overcome this challenge, we reformulate the problem using MPC. In the MPC framework, the long-horizon optimization is approximated by a sequence of finite-horizon subproblems. At each time step $t$, the UAV considers a prediction horizon of length $N_p$, defined by the index set $\mathcal{T}_t = \{t, t+1, \ldots, t+N_p\}$, where $\tau \in \mathcal{T}_t$ denotes the running time index within the prediction window. %Within the MPC formulation, channel prediction over the finite horizon must rely on a deterministic representation. While the large-scale LoS component can be computed from the predicted UAV–user geometry over the horizon $\mathcal{T}_t$, the small-scale fading term is stochastic and its future realizations are unknown. Accordingly, for future time steps within the prediction window, the channel is constructed using only the geometry-dependent LoS component, whereas the full Rician channel model is used at the currently executed time step $t$.
The resulting MPC problem at time step $t$ is
\begin{align}
\textbf{(P-MPC)}\quad
\max_{\substack{
\mathbf r(\tau),\mathbf v(\tau),\\
\mathbf w_n(\tau),\, \tau\in\mathcal T_t}}
& \sum_{\tau=t}^{t+N_p} \Bigg[
\psi_1\sum_{n=1}^{N} R_n(\tau) \nonumber\\
& - \psi_2 \|\mathbf r(\tau)-\mathbf r_B\|_2^2
- \psi_3 P_{\mathrm{prop}}(\mathbf{v}(\tau))
\Bigg] \nonumber\\
\text{s.t.}\quad
\text{C1:}\quad &\mathbf r(\tau+1)=\mathbf r(\tau)+\mathbf v(\tau)t_c,
\ \forall \tau \in \mathcal{T}_t, \nonumber\\
\text{C2a:}\quad & X_{\min} \le r_x(\tau) \le X_{\max},
\ \forall \tau \in \mathcal{T}_t, \nonumber\\
\text{C2b:}\quad & Y_{\min} \le r_y(\tau) \le Y_{\max},
\ \forall \tau \in \mathcal{T}_t, \nonumber\\
\text{C2c:}\quad & Z_{\min} \le r_z(\tau) \le Z_{\max},
\ \forall \tau \in \mathcal{T}_t, \nonumber\\
\text{C3:}\quad & \|\mathbf v_h(\tau)\|_2 \le V_{\max},
\ \forall \tau \in \mathcal{T}_t, \nonumber\\
\text{C4:}\quad & |v_z(\tau)| \le U_{\max},
\ \forall \tau \in \mathcal{T}_t, \nonumber\\
\text{C5:}\quad & \|\mathbf v(\tau+1)-\mathbf v(\tau)\|_2
\le a_{\max}t_c,
\ \forall \tau \in \mathcal{T}_t, \nonumber\\
\text{C6:}\quad & P_{\text{tot}}(\tau)\le P_{\max},
\ \forall \tau \in \mathcal{T}_t, \nonumber\\
\text{C7:}\quad & \frac{1}{\eta}\sum_{n=1}^{N}
\|\mathbf w_n(\tau)\|_2^2
\le P^{com}_{\max},
\ \forall \tau \in \mathcal{T}_t, \nonumber\\
\text{C8:}\quad & R_n(\tau)\ge R_n^{\min},
\ \forall n,\ \forall \tau \in \mathcal{T}_t.
\label{eq:20}
\end{align}

The constraints C1-C8 in problem (P-MPC) retain the same physical interpretations as those defined in (\textbf{P1}), with the time index $t$ replaced by $\tau \in \mathcal{T}_t$ over the prediction horizon. The joint trajectory and beamforming optimization problem under the MPC framework is inherently non-convex due to multiple factors. The SINR involves a ratio of quadratic terms in the beamforming vectors, making both the objective and rate constraints non-convex. The FBL achievable rate further introduces nonlinearity through the logarithmic and dispersion terms. In addition, the channel depends nonlinearly on the UAV trajectory via distance-dependent pathloss, while the propulsion power is a nonlinear function of UAV velocity. Although the motion constraints are convex, their combination with these nonlinearities renders the feasible set non-convex. Hence, the MPC-based problem cannot be solved in closed form and requires iterative methods such as alternating optimization (AO) and successive convex approximation (SCA) to obtain high-quality locally optimal solutions. For notational convenience, let $\Phi_t\!\left(\{\mathbf r(\tau),\mathbf v(\tau),\mathbf w_n(\tau)\}_{\tau\in\mathcal T_t}\right)$ denote the objective value of problem (\textbf{P-MPC}) evaluated over the prediction window $\mathcal T_t$.

To efficiently handle the non-convexity, we adopt an AO framework that decomposes the MPC problem into two coupled subproblems. At the beginning of each prediction window, a feasible UAV trajectory is initialized. Given the current trajectory iterate, the beamforming vectors are optimized by solving subproblem (\textbf{P2}), defined in Section III-A, while keeping the trajectory variables fixed. Subsequently, using the updated beamforming solution, the UAV trajectory and velocity are optimized via subproblem (\textbf{P3}), defined in Section III-B, with the beamformers held fixed. These two subproblems are solved alternately, where each block of variables is optimized with the other block fixed at its most recent iterate. The process is repeated until convergence within the MPC window.

\subsection{Beamforming Optimization Subproblem: Beamforming Design via Concave and Convex Surrogates}

For a given UAV trajectory $\{\mathbf r(\tau), \mathbf v(\tau)\}$, the optimization variables reduce to the per-user beamforming vectors $\{\mathbf w_n(\tau)\}$. 
Based on the achievable rate expression $R_n(\tau)$ defined in the system model, the per time step beamforming optimization problem is formulated as
\begin{equation}
\begin{aligned}
\textbf{(P2)}\quad
\max_{\{\mathbf w_n(\tau)\}}
& \sum_{\tau=t}^{t+N_p}\sum_{n=1}^{N} R_n(\tau) \\
\text{s.t.}\quad & \text{C7},\; \text{C8}.
\label{eq:21}
\end{aligned}
\end{equation}
%where the propulsion power $P_{\mathrm{prop}}(\mathbf v(\tau))$ is constant with respect to the beamforming vectors since the UAV trajectory is fixed in this subproblem.

To enhance convergence robustness of the SCA procedure in (\textbf{P2}),
a feasible beamforming initialization is generated at the beginning
of each MPC time step $t$.
Specifically, for each prediction index $\tau \in \mathcal T_t$,
we solve the following simplified max–min Shannon-rate problem:
\begin{equation}
\begin{aligned}
\textbf{(P2-Init)}\quad
\max_{\{\mathbf w_n(\tau)\}} 
& \min_{n} \log\!\left(1+\gamma_n(\tau)\right) \\
\text{s.t.}\quad & \text{C7}.
\label{eq:22}
\end{aligned}
\end{equation}
The resulting solution serves as the warm-start initialization for the subsequent SCA-based optimization of \textbf{(P2)}.

Problem \textbf{(P2)} is non-convex due to the quadratic fractional SINR terms in both the objective and QoS constraints, as well as the nonlinear FBL penalty embedded in $R_n(\tau)$. 
Accordingly, convexification is required. The UAV achievable rate under the FBL regime, $R_n(\tau)=C_n(\tau)-D_n(\tau)$, is inherently non-convex due to the coupling between the Shannon term $C_n(\tau)$ and the FBL penalty term $D_n(\tau)$. Accordingly, SCA is employed.

\paragraph{Concave Surrogate of the Shannon Term}

The Shannon term can be expressed as
\begin{align}
C_n(\tau) &= \log\big(1+\gamma_{n}(\tau)\big) \nonumber \\
&=
\log\!\big(S_n(\tau)+I_n(\tau)\big)
-
\log\!\big(I_n(\tau)\big),
\label{log_functions}
\end{align}
where
$S_n(\tau)=|\mathbf h_n^H(\tau)\mathbf w_n(\tau)|^2$
and
$I_n(\tau)=\sum_{k\neq n}|\mathbf h_n^H(\tau)\mathbf w_k(\tau)|^2+\sigma^2$.
At SCA iteration $i$, by applying the first-order Taylor expansion of the concave logarithmic function to the Shannon term in Eq.~(\ref{log_functions}) at the point $(S_n^{(i)}(\tau), I_n^{(i)}(\tau))$, we obtain the following global concave lower bound for the Shannon term, as shown in Eq.~(\ref{eq:Shannon_main}). The detailed derivation is provided in Appendix~A.
\begin{equation}
C_n(\tau)
\ge
C_n^{(i)}(\tau)
+
L_n^{(i)}(\tau)
\triangleq
\overline C_n^{(i)}(\tau),
\label{eq:Shannon_main}
\end{equation}
with
\begin{equation}
C_n^{(i)}(\tau)
=
\log\!\left(1+\gamma_n^{(i)}(\tau)\right),
\end{equation}
and where the increment term is given by
\begin{align}
L_n^{(i)}(\tau)
&=
\frac{2\Re\!\left\{
\big(\chi_n^{(i)}(\tau)\big)^*
\mathbf h_n^H(\tau)\mathbf w_n(\tau)
\right\}}
{I_n^{(i)}(\tau)}
-
\gamma_n^{(i)}(\tau)
\nonumber\\
&\quad
-
\eta_n^{(i)}(\tau)\big(S_n(\tau)+I_n(\tau)\big),
\label{eq:L_main_TVT}
\end{align}
where
\begin{equation}
\eta_n^{(i)}(\tau)
=
\frac{S_n^{(i)}(\tau)}
{I_n^{(i)}(\tau)\big(S_n^{(i)}(\tau)+I_n^{(i)}(\tau)\big)},
\label{eq:eta_def}
\end{equation}
\(
\chi_n^{(i)}(\tau)=\mathbf h_n^H(\tau)\mathbf w_n^{(i)}(\tau)
\)
and
\(
\gamma_n^{(i)}(\tau)=\frac{S_n^{(i)}(\tau)}{I_n^{(i)}(\tau)}.
\) The detailed derivation from the Taylor expansion to the affine expression of $L_n^{(i)}(\tau)$ is provided in Appendix~A.

The quadratic signal term is linearized as
\begin{align}
S_n(\tau)
&=
|\mathbf h_n^H(\tau)\mathbf w_n(\tau)|^2
\nonumber\\
&\ge
2\Re\!\left\{(\chi_n^{(i)}(\tau))^*
\,\mathbf h_n^H(\tau)\mathbf w_n(\tau)\right\}
-
|\chi_n^{(i)}(\tau)|^2 .
\label{eq:signal_lb_TVT}
\end{align}
which ensures that $\overline C_n^{(i)}(\tau)$
is concave in the beamforming variables. The validity of
this linearization is ensured by a trust-region condition
that maintains the positivity of the approximated signal
term, as stated in Appendix~A.
%%%%%%%%%%%%%%%%%%%%%%%%%%%%%%%%%%%%%%%%%%%%%%%%%%%%%
\paragraph{Convex Surrogate of the FBL Penalty}
The FBL penalty term is given by
\begin{equation}
D_n(\tau)=c_n\sqrt{V_n(\tau)},
\label{eq:D_main}
\end{equation}
where $c_n=\frac{Q^{-1}(\epsilon_n)}{\sqrt{L}}$ is a constant determined by the target decoding error probability $\epsilon_n$ and the blocklength $L$, and
\begin{equation}
V_n(\tau)=1-\frac{1}{(1+\gamma_n(\tau))^2}=1 - \frac{I_n(\tau)^2}{\left(S_n(\tau)+I_n(\tau)\right)^2}.
\label{eq:V_main}
\end{equation}
Since $\sqrt{x}$ is concave for $x>0$, its first-order Taylor expansion
at $V_n^{(i)}(\tau)$ yields the global upper bound
\begin{equation}
\sqrt{V_n(\tau)}
\le
B_n^{(i)}(\tau)
-
A_n^{(i)}(\tau)
\frac{I_n(\tau)^2}{\big(S_n(\tau)+I_n(\tau)\big)^2},
\label{eq:sqrt_bound_main}
\end{equation}
where
\begin{align}
A_n^{(i)}(\tau)
&=\frac{1}{2\sqrt{V_n^{(i)}(\tau)}},\\
B_n^{(i)}(\tau)
&=A_n^{(i)}(\tau)
+\frac{1}{2}\sqrt{V_n^{(i)}(\tau)}.
\end{align}
The detailed derivation of Eq.~\eqref{eq:sqrt_bound_main} is provided in Appendix~\ref{appendix:fbl}.

The fractional term
\begin{equation}
\frac{I_n(\tau)^2}{\big(S_n(\tau)+I_n(\tau)\big)^2}
\label{eq:fraction_main}
\end{equation}
appearing in Eq.~\eqref{eq:sqrt_bound_main} admits the affine upper bound
\begin{align}
\frac{I_n(\tau)^2}{\big(S_n(\tau)+I_n(\tau)\big)^2}
&\le
\alpha_n^{(i)}(\tau) I_n(\tau)
+
\beta_n^{(i)}(\tau) \big(S_n(\tau)+I_n(\tau)\big) \nonumber\\
&+
\psi_n^{(i)}(\tau),
\label{eq:fraction_affine_main}
\end{align}
with
\begin{align}
\alpha_n^{(i)}(\tau)
&=\frac{4 I_n^{(i)}(\tau)}
{\big(S_n^{(i)}(\tau)+I_n^{(i)}(\tau)\big)^2},\\
\beta_n^{(i)}(\tau)
&=\frac{2\big(I_n^{(i)}(\tau)\big)^2}
{\big(S_n^{(i)}(\tau)+I_n^{(i)}(\tau)\big)^3},\\
\psi_n^{(i)}(\tau)
&=\frac{\big(I_n^{(i)}(\tau)\big)^2}
{\big(S_n^{(i)}(\tau)+I_n^{(i)}(\tau)\big)^2}.
\end{align}
The detailed derivation of
Eq.~\eqref{eq:fraction_affine_main} including the associated trust-region condition,
is provided in Appendix~\ref{appendix:fbl}. To obtain an affine surrogate in the beamforming variables,
the signal and interference terms are approximated by their
first-order affine lower bounds at the current SCA iterate:
\begin{align}
\bar{S}_n(\tau)
&=
2\Re \!\left\{
\big(\mathbf{h}_n^H(\tau)\mathbf{w}_n^{(i)}(\tau)\big)^*
\mathbf{h}_n^H(\tau)\mathbf{w}_n(\tau)
\right\} \nonumber \\
&-
\big|\mathbf{h}_n^H(\tau)\mathbf{w}_n^{(i)}(\tau)\big|^2,
\\
\bar{I}_n(\tau)
&=
\sum_{k\neq n}
\Big[
2\Re \!\left\{
\big(\mathbf{h}_n^H(\tau)\mathbf{w}_k^{(i)}(\tau)\big)^*
\mathbf{h}_n^H(\tau)\mathbf{w}_k(\tau)
\right\} \nonumber \\
&-
\big|\mathbf{h}_n^H(\tau)\mathbf{w}_k^{(i)}(\tau)\big|^2
\Big]
+ \sigma^2 .
\end{align}

Replacing $S_n(\tau)$ and $I_n(\tau)$ in
Eq.~\eqref{eq:fraction_affine_main}
by $\bar S_n(\tau)$ and $\bar I_n(\tau)$,
and substituting Eq.~\eqref{eq:fraction_affine_main}
into Eq.~\eqref{eq:sqrt_bound_main},
yields
\begin{equation}
D_n(\tau)
\le
\overline D_n^{(i)}(\tau),
\label{eq:Dbar_main}
\end{equation}
where
\begin{align}
\overline D_n^{(i)}(\tau)=
c_n(\tau)
\Big[
B_n^{(i)}(\tau)
- A_n^{(i)}(\tau)\,
\Xi_n^{(i)}(\tau)
\Big]
\label{eq:Dbar_explicit}
\end{align}
where the affine term $\Xi_n^{(i)}(\tau)$ is defined as
\begin{equation}
\Xi_n^{(i)}(\tau)
=
\alpha_n^{(i)}(\tau)\,\bar{I}_n(\tau)
+ \beta_n^{(i)}(\tau)\big(\bar{S}_n(\tau)+\bar{I}_n(\tau)\big)
+ \psi_n^{(i)}(\tau),
\label{eq:Xi_definition}
\end{equation}

%%%%%%%%%%%%%%%%%%%%%%%%%%%%%%%%%%%%%%%%%%%%%%%%
Combining the two bounds, the achievable rate is approximated as
\begin{equation}
R_n^{(i)}(\tau)
=
\overline{C}_n^{(i)}(\tau)
-
\overline{D}_n^{(i)}(\tau),
\label{eq:final_surrogate}
\end{equation}
which preserves feasibility and enables convex optimization at each SCA iteration. Accordingly, at SCA iteration $i$, Problem \textbf{(P2)} is approximated by the following convex problem:
\begin{align}
\textbf{(P2)} \quad
&\max_{\{w_n(\tau)\}} 
 \sum_{\tau=t}^{t+N_p} \sum_{n=1}^{N}
\Big(
\overline{C}_n^{(i)}(\tau)
-
\overline{D}_n^{(i)}(\tau)
\Big) \nonumber
\\
\text{s.t.} \quad
& \text{C7: } \frac{1}{\eta}
\sum_{n=1}^{N} \|w_n(\tau)\|_2^2
\le P_{\max}^{\text{com}},
\quad \forall \tau \in \mathcal{T}_t, \nonumber
\\
& \text{C8: } 
\overline{C}_n^{(i)}(\tau)
-
\overline{D}_n^{(i)}(\tau)
\ge R_n^{\min},
\quad \forall n,\ \forall \tau \in \mathcal{T}_t.
\label{eq:27}
\end{align}

\subsection{Trajectory Optimization Subproblem: UAV Path and Velocity Update}
For the trajectory optimization, the beamforming vectors 
$\{\mathbf{w}_n^{(i+1)}(\tau)\}$ obtained from the preceding subproblem are fixed. 
The objective is to optimize the UAV’s position and velocity profiles 
$\{\mathbf{r}(\tau), \mathbf{v}(\tau)\}_{\tau \in \mathcal{T}_t}$ within the current prediction horizon 
$\mathcal{T}_t = \{t, t+1, \ldots, t+N_p\}$ to further enhance the system performance 
while satisfying the UAV’s motion and power constraints. 
The resulting optimization problem is formulated as
\begin{equation}
\begin{aligned}
\textbf{(P3)}\quad
\max_{\{\mathbf r(\tau),\mathbf v(\tau)\}}
& \sum_{\tau=t}^{t+N_p} \Bigg[
\psi_1\sum_{n=1}^{N} R_n(\tau) \\
&\quad - \psi_2 \|\mathbf r(\tau)-\mathbf r_B\|_2^2
- \psi_3 P_{\mathrm{prop}}(\mathbf{v}(\tau))
\Bigg] \\
\text{s.t.}\quad
& \text{C1--C6},\ \text{C8}.
\label{eq:P3}
\end{aligned}
\end{equation}

Problem \textbf{P3} in Eq.~(\ref{eq:P3}) is non-convex due to the nonlinear dependence of the achievable rate on 
the UAV position through the distance-dependent pathloss 
$\beta_n(\tau) = B_0 d_n(\tau)^{-\rho}$, where 
$d_n(\tau) = \|\mathbf{u}_n - \mathbf{r}(\tau)\|_2$, 
and due to the non-convex propulsion power $P_{\mathrm{prop}}(v(\tau))$, which appears in both the objective function and constraint C6. The convex reformulation of the propulsion power used in (\textbf{P3})  is provided in Appendix~C.
To handle the remaining non-convexities, SCA is applied to construct convex surrogate functions that are tight at the current iterate and satisfy global bounding conditions.

For efficient solution, it is important to reformulate the channel model into a tractable form. Therefore, the expression in Eq.~(\ref{eq:12}) can be rewritten for this subproblem as
\begin{equation}
\mathbf{h}_n(\tau)
=
\sqrt{\beta_n(\tau)}\,\hat{\mathbf{h}}_n(\tau),
\label{eq:30}
\end{equation}

\begin{equation}
\hat{\mathbf{h}}_n(\tau)
=
\sqrt{\frac{K_n}{K_n + 1}}\,\mathbf{a}(\theta_n(\tau))
+
\sqrt{\frac{1}{K_n + 1}}\,\mathbf{g}_n(\tau).
\label{eq:31}
\end{equation}
Substituting Eq.~(\ref{eq:30}) into the received signal model in Eq.~(\ref{eq:15}), the instantaneous SINR of user $n$ can be explicitly written as a function of the UAV location:
\begin{equation}
\gamma_n(\tau;\mathbf{r}) =
\frac{
B_0\,\|\mathbf{u}_n - \mathbf{r}(\tau)\|^{-\rho}
\big|\hat{\mathbf{h}}_n^{H}(\tau)\mathbf{w}_n(\tau)\big|^2
}{
\sum_{k\neq n}B_0\,\|\mathbf{u}_n - \mathbf{r}(\tau)\|^{-\rho}\big|\hat{\mathbf{h}}_n^{H}(\tau)\mathbf{w}_k(\tau)\big|^2
+\sigma^2 }.
\label{eq:25}
\end{equation}
To simplify notation, define
\begin{equation}
\begin{aligned}
A_n(\tau) 
&= \big|\hat{\mathbf{h}}_n^{H}(\tau)\mathbf{w}_n(\tau)\big|^2, \\
B_n(\tau) 
&= \sum_{k\neq n}
\big|\hat{\mathbf{h}}_n^{H}(\tau)\mathbf{w}_k(\tau)\big|^2, \\
\kappa 
&= \frac{\sigma^2}{B_0}.
\end{aligned}
\label{eq:AnBn}
\end{equation}
Then, the SINR can be expressed as
\begin{equation}
\begin{aligned}
\gamma_n(\tau; \mathbf{r})
&= \frac{A_n(\tau)}
{B_n(\tau) + \kappa d_n(\tau;\mathbf{r})^{\rho}}, \\
d_n(\tau; \mathbf{r})
&= \|\mathbf{u}_n - \mathbf{r}(\tau)\|_2.
\end{aligned}
\label{eq:gamma_dn}
\end{equation}
The corresponding FBL achievable rate is
\begin{equation}
\begin{aligned}
R_n(\tau;\mathbf{r})
&= \underbrace{
\log\!\big(1+\gamma_n(\tau;\mathbf{r})\big)
}_{C_n(\tau;\mathbf{r})}  \\
&\quad - \underbrace{
c_n\,\sqrt{\,1-\big(1+\gamma_n(\tau;\mathbf{r})\big)^{-2}\,}
}_{D_n(\tau;\mathbf{r})}.
\end{aligned}
\label{eq:26}
\end{equation}
where $c_n=\frac{Q^{-1}(\varepsilon)}{\sqrt{L}}$ (with target error probability $\varepsilon$ and blocklength $L$), and $\gamma_n(\tau;\mathbf{r})$ is given in Eq.~(\ref{eq:25}). The Shannon term in Eq.~(\ref{eq:26}) depends on the UAV position only through the distance
\begin{equation}
d_n(\tau;\mathbf{r})=\|\mathbf{u}_n-\mathbf{r}(\tau)\|_2.
\label{eq:28}
\end{equation}
For convex reformulation within the SCA framework, we introduce auxiliary distance variables $\tilde d_n(\tau)$ and replace the distance expression $d_n(\tau;\mathbf r)=\|\mathbf u_n-\mathbf r(\tau)\|_2$ by the second-order cone constraint
\begin{equation}
\|\mathbf u_n - \mathbf r(\tau)\|_2 \le \tilde d_n(\tau),
\quad \forall n,\ \forall \tau \in \mathcal T_t .
\label{eq:aux_distance}
\end{equation}
To confine the auxiliary distance to the verified concavity interval, we further impose explicit bounds on $\tilde d_n(\tau)$ as
\begin{equation}
d_{\min} \le \tilde d_n(\tau) \le d_{\max},
\quad \forall n,\ \forall \tau \in \mathcal T_t ,
\label{eq:distance_bounds}
\end{equation}
Accordingly, we rewrite the SINR as a function of the auxiliary distance variable
\begin{equation}
\gamma_n(\tau)=\frac{A_n(\tau)}{B_n(\tau)+\kappa\,\tilde d_n(\tau)^{\rho}},
\label{eq:gamma_tilde}
\end{equation}
where the dependence on the trajectory $\mathbf r(\tau)$ is captured through Eq.~\eqref{eq:aux_distance}.
Since the achievable rate is monotonically decreasing with respect to
the UAV--user distance, the inequality becomes tight at the optimum.
Consequently, $C_n(\tau)=\log(1+\gamma_n(\tau))$ is a convex and monotonically decreasing function of $\tilde d_n(\tau)$.
Therefore, its first-order Taylor expansion around the previous iterate $\tilde d_n^{(i)}(\tau)$ yields a global affine lower bound.
\begin{equation}
\begin{aligned}
C_n(\tau)
&\ge C_n^{(i)}(\tau)
+
\left.
\frac{\partial C_n(\tau)}
{\partial \tilde d_n(\tau)}
\right|_{\tilde d_n^{(i)}(\tau)}
\big(
\tilde d_n(\tau)-\tilde d_n^{(i)}(\tau)
\big).
\end{aligned}
\label{eq:29}
\end{equation}
where
\begin{equation}
C_n^{(i)}(\tau)
= \log\!\left(1 + \frac{A_n(\tau)}{B_n(\tau) + \kappa (\tilde d_n^{(i)}(\tau))^{\rho}}\right).
\label{eq:Cn_i}
\end{equation}
and
\begin{equation}
\begin{aligned}
\left.
\frac{\partial C_n(\tau)}
{\partial \tilde d_n(\tau)}
\right|_{\tilde d_n^{(i)}(\tau)}
&= - \frac{
A_n(\tau)\,\kappa\,\rho
\left(\tilde d_n^{(i)}(\tau)\right)^{\rho - 1}
}{
\left(1 + \gamma_n^{(i)}(\tau)\right)
\left(
B_n(\tau)
+ \kappa
\left(\tilde d_n^{(i)}(\tau)\right)^{\rho}
\right)^2
}.
\end{aligned}
\label{eq:dCn_ddn}
\end{equation}
with
\begin{equation}
\gamma_n^{(i)}(\tau) = \frac{A_n(\tau)}{B_n(\tau) + \kappa (\tilde d_n^{(i)}(\tau))^{\rho}}.
\label{eq:gamma_i}
\end{equation}
Finally, combining the affine lower bound Eq.~\eqref{eq:29} with the SOC coupling constraint Eq.~\eqref{eq:aux_distance} yields a convex approximation of $C_n(\tau)$ in each SCA iteration.

The finite-blocklength (FBL) penalty term is given by 
$D_n(\tau)=c_n\sqrt{1-(1+\gamma_n(\tau))^{-2}}$.
The dependence on the UAV position $\mathbf r(\tau)$ is captured through the second-order cone constraint 
$\|\mathbf u_n-\mathbf r(\tau)\|_2\le \tilde d_n(\tau)$. 
The function $D_n(\tau)$ is differentiable with respect to the auxiliary distance
variable $\tilde d_n(\tau)$ through the SINR expression $\gamma_n(\tau)$.
Let $\tilde d_n^{(i)}(\tau)$ denote the auxiliary distance at the current SCA iterate.
 By the chain rule,
\begin{equation}
\begin{aligned}
\left.
\frac{\partial D_n(\tau)}
{\partial \tilde d_n(\tau)}
\right|_{\tilde d_n^{(i)}(\tau)}
&=
\left.
\frac{\partial D_n(\tau)}
{\partial \gamma_n(\tau)}
\right|_{\gamma_n^{(i)}(\tau)}
\cdot
\left.
\frac{\partial \gamma_n(\tau)}
{\partial \tilde d_n(\tau)}
\right|_{\tilde d_n^{(i)}(\tau)} .
\end{aligned}
\label{eq:dDdd_chain}
\end{equation}
The first term in Eq.~\eqref{eq:dDdd_chain} follows from $D_n(\tau)$
\begin{equation}
\left.\frac{\partial D_n(\tau)}{\partial \gamma_n(\tau)}\right|_{\gamma_n^{(i)}(\tau)}
=
c_n\,
\frac{\left(1+\gamma_n^{(i)}(\tau)\right)^{-3}}
{\sqrt{\,1-\left(1+\gamma_n^{(i)}(\tau)\right)^{-2}\,}} .
\label{eq:dDdgamma}
\end{equation}
From the SINR expression in Eq.~\eqref{eq:gamma_tilde}, we obtain
\begin{equation}
\left.\frac{\partial \gamma_n(\tau)}{\partial \tilde d_n(\tau)}\right|_{\tilde d_n^{(i)}(\tau)}
=
-
\frac{
A_n(\tau)\,\kappa\,\rho
\left(\tilde d_n^{(i)}(\tau)\right)^{\rho-1}
}{
\left(B_n(\tau) + \kappa\left(\tilde d_n^{(i)}(\tau)\right)^{\rho}\right)^2 } .
\label{eq:dgd_d}
\end{equation}

Substituting Eq.~\eqref{eq:dDdgamma}–\eqref{eq:dgd_d} into Eq.~\eqref{eq:dDdd_chain} yields the local slope 

\begin{equation}
g_{D,n}^{(i)}(\tau)
=
\left.\frac{\partial D_n(\tau)}{\partial \tilde d_n(\tau)}\right|_{\tilde d_n^{(i)}(\tau)} .
\label{eq:gD_def}
\end{equation}

\noindent\textbf{Remark (Strict concavity of the dispersion penalty on the feasible distance interval):}
The dispersion penalty $D_n(\tau)=c_n\sqrt{1-(1+\gamma_n(\tau))^{-2}}$ depends on the UAV trajectory only through the auxiliary distance $\tilde d_n(\tau)$ in Eq.~\eqref{eq:aux_distance} via $\gamma_n(\tau)$ in Eq.~\eqref{eq:gamma_tilde}. By differentiating Eq.~\eqref{eq:dDdgamma}--\eqref{eq:dgd_d} and applying the chain rule, $D_n(d)$ is twice continuously differentiable for $d>0$ and admits a well-defined second derivative $D_n''(d)$. For the considered corridor-based scenario, we enforce $d\in[d_{\min},d_{\max}]$ with $d_{\min}>0$ a small positive constant introduced to avoid the singularity at $d=0$ and $d_{\max}=2000$~m. As shown in Fig.~2, $D_n''(d)<0$ over $[d_{\min},d_{\max}]$ (and we verified this for all users, time slots, and AO iterations), hence $D_n(\tau)$ is \emph{strictly concave} in $\tilde d_n(\tau)$ on the feasible set. Therefore, the first-order Taylor expansion at $\tilde d_n^{(i)}(\tau)$ yields the valid affine upper bound in Eq.~\eqref{eq:D_upper_bound}, which is tight at $\tilde d_n^{(i)}(\tau)$ and, together with Eq.~\eqref{eq:aux_distance}, leads to a convex SCA surrogate.

\begin{equation}
D_n(\tau)
\;\le\;
D_n\!\left(\tilde d_n^{(i)}(\tau)\right)
+
g_{D,n}^{(i)}(\tau)\,
\big( \tilde d_n(\tau) - \tilde d_n^{(i)}(\tau) \big) ,
\label{eq:D_upper_bound}
\end{equation}
The upper bound in Eq.~\eqref{eq:D_upper_bound} is affine in $\tilde d_n(\tau)$ and tight at the linearization point. Together with the SOC constraint in Eq.~\eqref{eq:aux_distance}, it yields a convex approximation that can be efficiently handled in the SCA-based trajectory optimization. 

\begin{figure}[H]
    \centering
    \includegraphics[width= 0.90\linewidth]{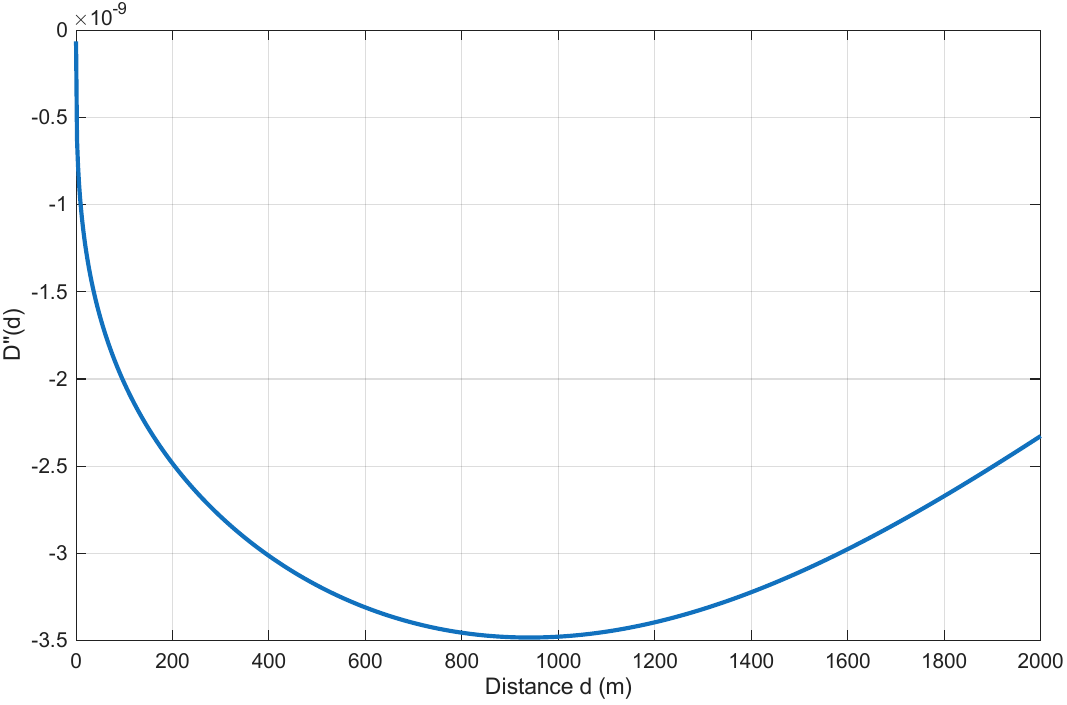}
    \caption{Second derivative $D_n''(d)$ of the FBL dispersion term.}
    \label{fig:49}
\end{figure}

By combining the affine lower bound of the Shannon term and the affine upper bound of the FBL penalty with the SOC coupling constraint Eq.~\eqref{eq:aux_distance}, the auxiliary distance bounds Eq.~\eqref{eq:distance_bounds}, and the convex surrogate of the propulsion power $P_{\mathrm{prop}}^{\mathrm{ub}}(\mathbf v(\tau))$, problem (\textbf{P3}) is transformed into a convex optimization problem at each SCA iteration. Consequently, the trajectory and velocity variables can be efficiently updated using standard convex optimization solvers within the receding-horizon MPC framework.

To preserve recursive feasibility of the receding-horizon MPC scheme, if either the beamforming or trajectory convex subproblem becomes infeasible during an AO iteration due to strict QoS constraint, the most recent feasible solution is retained and used for system update. The overall MPC-based joint trajectory and beamforming optimization procedure is summarized in Algorithm~1. 

\subsection{Convergence Analysis}
In each MPC window, the proposed algorithm adopts an AO framework combined with SCA. 
Let $\Phi_t(\mathbf{w},\mathbf{r},\mathbf{v})$ denote the objective function at time step $t$, and let 
$\Phi_t^{(i)}$ be the objective value at the $i$-th AO iteration.

At iteration $i$, given $(\mathbf{r}^{(i)},\mathbf{v}^{(i)})$, 
the beamforming variables are updated by solving the convexified subproblem. 
Since the SCA surrogate is tight at the current point and preserves first-order optimality, we obtain
\begin{equation}
\Phi_t(\mathbf{w}^{(i+1)}, \mathbf{r}^{(i)}, \mathbf{v}^{(i)})
\ge
\Phi_t(\mathbf{w}^{(i)}, \mathbf{r}^{(i)}, \mathbf{v}^{(i)}).
\label{eq:bf_monotonic}
\end{equation}

Next, fixing $\mathbf{w}^{(i+1)}$, the trajectory and velocity variables are updated via convex approximation. 
By the tightness property of SCA, the following inequality holds:
\begin{equation}
\Phi_t(\mathbf{w}^{(i+1)}, \mathbf{r}^{(i+1)}, \mathbf{v}^{(i+1)})
\ge
\Phi_t(\mathbf{w}^{(i+1)}, \mathbf{r}^{(i)}, \mathbf{v}^{(i)}).
\label{eq:traj_monotonic}
\end{equation}

Combining Eq.~\eqref{eq:bf_monotonic} and Eq.~\eqref{eq:traj_monotonic}, we have
\begin{equation}
\Phi_t^{(i+1)} \ge \Phi_t^{(i)},
\label{eq:overall_monotonic}
\end{equation}
which shows that the objective sequence $\{\Phi_t^{(i)}\}$ is monotonically non-decreasing.

Moreover, since the feasible set defined by the transmit power, propulsion power, and QoS constraints is compact and the objective function is continuous, the objective value is upper bounded, i.e.,
\begin{equation}
\Phi_t^{(i)} \le \Phi_t^{\max} < \infty.
\label{eq:boundedness}
\end{equation}

Therefore, the sequence $\{\Phi_t^{(i)}\}$ is monotonically non-decreasing and upper bounded, and thus converges. 
Under successful solution of each convex subproblem, the proposed algorithm 1 converges to a stationary point of the original non-convex problem in each MPC window.
\begin{algorithm}[t]
\caption{MPC-Based Joint Trajectory and Beamforming Optimization}
\label{alg:mpc_final}
\small
\begin{algorithmic}[1]
\Require Initial UAV state $(\mathbf r(1)=\mathbf r_A,\mathbf v(1))$, 
destination $\mathbf r_B$, acceptable deviation $\varepsilon$, 
initial beamformers $\{\mathbf w_n(1)\}$,
prediction horizon $N_p$, 
maximum AO iterations $i_{\max}$.

\State $t \gets 1$

\While{$\|\mathbf r(t)-\mathbf r_B\|_2 > \varepsilon$}

    \State Define prediction window 
$\mathcal{T}_t = \{t, t+1, \dots, t+N_p-1\}$
    \State Initialize feasible trajectory and beamformers over $\mathcal{T}_t$
    \State $i \gets 0$

    \Repeat
        \State Solve convexified beamforming subproblem (\textbf{P2})
        \State Solve convexified trajectory subproblem (\textbf{P3})
        \State $i \gets i + 1$
    \Until{$i \ge i_{\max}$}

    \State \textbf{Receding-horizon implementation:}
    \State \hspace{1em} Extract optimized sequences 
    $\{\mathbf r^{\star}(\tau),\mathbf v^{\star}(\tau),\mathbf w_n^{\star}(\tau)\}_{\tau\in\mathcal T_t}$
    \State \hspace{1em} Apply first-step control:
    \State \hspace{2em} $\mathbf v(t) \gets \mathbf v^{\star}(t)$
    \State \hspace{2em} $\mathbf w_n(t) \gets \mathbf w_n^{\star}(t),\ \forall n$
    \State \hspace{2em} $\mathbf r(t{+}1) \gets \mathbf r(t) + \mathbf v^{\star}(t)t_c + \text{disturbance}(t)$

    \State $t \gets t + 1$

\EndWhile

\State \textbf{Output:} UAV trajectory, velocity, and beamforming vectors.

\end{algorithmic}
\end{algorithm}
\subsection{Computational Complexity Analysis}

The computational complexity of the proposed MPC-based algorithm 1 is determined by the convex subproblems solved within each AO iteration at every MPC step. At time step $t$, the optimization is performed over a prediction horizon of length $N_p$, and the AO procedure is executed for at most $i_{\max}$ iterations.

In each AO iteration, one beamforming subproblem and one trajectory subproblem are solved sequentially. For the beamforming update, the decision variables are the beamforming vectors $\{\mathbf{w}_n(\tau)\}$ for $n=1,\ldots,N$ and $\tau\in\mathcal{T}_t$, where $\mathbf{w}_n(\tau)\in\mathbb{C}^{M}$. The number of optimization variables scales as $\mathcal{O}(MNN_p)$, while the number of constraints (including per-user QoS and per-time-step transmit power constraints) scales as $\mathcal{O}(NN_p)$. Solving the resulting convex program via an interior-point method (IPM), whose per-iteration cost depends on both variables and constraints, yields a worst-case complexity on the order of
\begin{equation}
\mathcal{C}_{\mathrm{bf}}
=
\mathcal{O}\!\left(\left(MNN_p+NN_p\right)^3\right)
=
\mathcal{O}\!\left(\left((MN+N)N_p\right)^3\right).
\end{equation}

For the trajectory update, the UAV position and velocity variables over the horizon contribute $6N_p$ real variables, and the SCA reformulation introduces user-dependent auxiliary variables per time step, resulting in $\mathcal{O}((N+6)N_p)$ optimization variables. The associated motion, power, SOC, and per-user rate constraints scale as $\mathcal{O}((N+1)N_p)$. Hence, the corresponding convex subproblem has worst-case complexity
\begin{equation}
\mathcal{C}_{\mathrm{tr}}
=
\mathcal{O}\!\left(\left((N+6)N_p+(N+1)N_p\right)^3\right)
=
\mathcal{O}\!\left(\left((2N+7)N_p\right)^3\right).
\end{equation}

Since each AO iteration solves both subproblems once, the per-iteration computational cost is
\begin{equation}
\mathcal{C}_{\mathrm{AO}}
=
\mathcal{O}\!\left(
\left((MN+N)N_p\right)^3
+
\left((2N+7)N_p\right)^3
\right).
\end{equation}

Considering at most $i_{\max}$ AO iterations per MPC window and $T$ MPC steps over the mission duration, the overall worst-case computational complexity is given by
\begin{equation}
\mathcal{C}_{\mathrm{total}}
=
\mathcal{O}\!\left(
T\, i_{\max}
\left[
\left((MN+N)N_p\right)^3
+
\left((2N+7)N_p\right)^3
\right]\right).
\end{equation}
Therefore, the proposed algorithm exhibits polynomial-time complexity and remains computationally tractable for moderate values of $M$, $N$, and $N_p$.

\section{RESULTS AND DISCUSSION}
In this section, MATLAB simulations are conducted to evaluate the performance of the proposed online MPC framework under FBL URLLC constraints. All optimization problems are solved using convex programming tools with the MOSEK solver. Unless otherwise stated, all system parameters follow Table~\ref{tab:simulation_parameters}.

\begin{table}[t]
\caption{Simulation Parameters}
\label{tab:simulation_parameters}
\centering
\begin{tabular}{l c}
\hline
\textbf{Parameter} & \textbf{Value} \\
\hline
Number of users $N$ & $3$ \\
Number of UAV antennas $M$ & $4$ \\
User deployment & Corridor along $r_A \rightarrow r_B$ \\
Corridor width & $200$ m \\
User height & Uniform in $[0.5, 3]$ m \\
UAV initial position $r_A$ & $[750,\,50,\,500]$ m \\
UAV destination $r_B$ & $[1000,\,200,\,300]$ m \\
Time step $t_c$ & $1$ s \\
Prediction horizon $N_p$ & $5$ \\
Maximum horizontal speed $V_{\max}$ & $15$ m/s \\
Maximum vertical speed $U_{\max}$ & $10$ m/s \\
Maximum acceleration $a_{\max}$ & $4$ m/s$^2$ \\
Altitude limits & $[100,\,900]$ m \\
Horizontal flight region & $[-1500,1500]$ m \\
System bandwidth & $5$ MHz \\
Noise PSD & $-174$ dBm/Hz \\
Pathloss exponent $\rho$ & $2.3$ \\
Rician factor $K_R$ & $6$ dB \\
Blocklength $L$ & $1000$ \\
Target error probability $\epsilon$ & $10^{-5}$ \\
RF transmit power budget $P^{com}_{max}$ & $30$ dBm (1 W) \\
Maximum total power budget & $230$ W \\
Power amplifier efficiency $\eta$ & $0.5$ \\
UAV weight $W$ & $39.2$ N \\
Air density & $1.225$ kg/m$^3$ \\
Rotor disk area $S$ & $0.503$ m$^2$ \\
Blade profile coefficient $\zeta$ & $0.08$ \\
\hline
\end{tabular}
\end{table}

\begin{figure}[t]
% -------- First Row --------
\begin{minipage}{0.45\textwidth}
    \centering
    \includegraphics[width=0.95\linewidth]{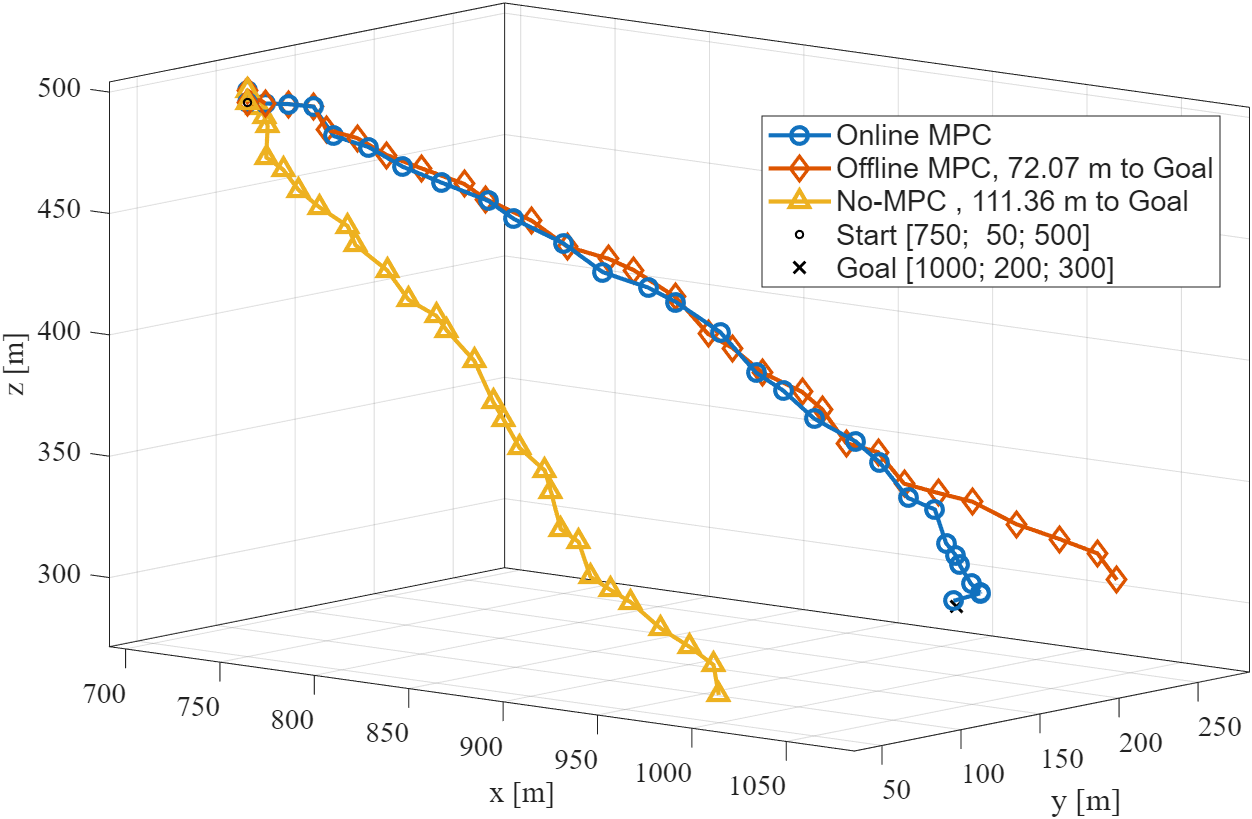}
    \captionof{figure}{UAV trajectories under online MPC and benchmark schemes.}
    \label{fig:trajectory_comparison}
\end{minipage}
\hfill
\begin{minipage}{0.45\textwidth}
    \centering
    \includegraphics[width=0.9\linewidth]{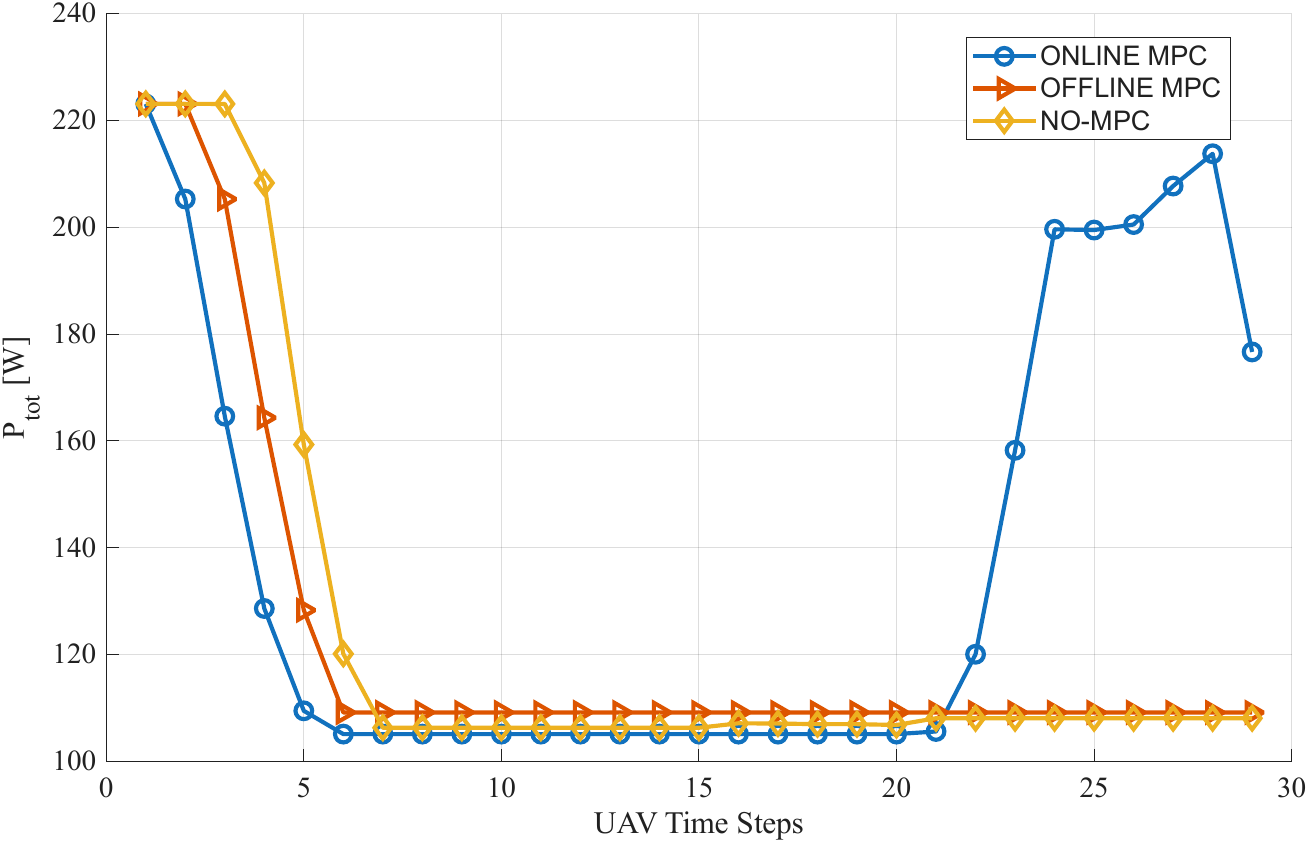}
    \captionof{figure}{Total instantaneous power under bounded disturbance of 6 m.}
    \label{fig:total_power_dev6}
\end{minipage}
\end{figure}

Fig.~\ref{fig:trajectory_comparison} compares the UAV trajectories under the proposed online MPC and the benchmark schemes over the same mission duration $T$ in the presence of a bounded additive position disturbance of magnitude $6$\,m at each time step. 
Under identical time-step constraints, the proposed online MPC successfully reaches the prescribed destination, whereas the offline MPC and non-MPC schemes remain $72.07$\,m and $111.36$\,m away from the target, respectively. 
The significant residual errors in the open-loop schemes result from disturbance accumulation without state-feedback correction. 
In contrast, the receding-horizon re-optimization mechanism compensates for trajectory deviations at each time step, ensuring accurate convergence to the designated destination within the allotted mission time. 
These results confirm the robustness advantage of the proposed closed-loop design under bounded execution uncertainties.

Fig.~\ref{fig:total_power_dev6}  shows the instantaneous total power consumption under the same disturbance setting. 
All schemes exhibit elevated power in the initial UAV time steps due to acceleration and trajectory alignment. However, the proposed online MPC demonstrates adaptive power adjustments in later stages to compensate for accumulated deviations, whereas the open-loop schemes maintain nearly constant power profiles without corrective action. This adaptive behavior enables accurate convergence while preserving practical power efficiency.

\begin{figure*}[t]
\centering
% -------- First Row --------
\begin{minipage}{0.45\textwidth}
    \centering
    \includegraphics[width=0.9\linewidth]{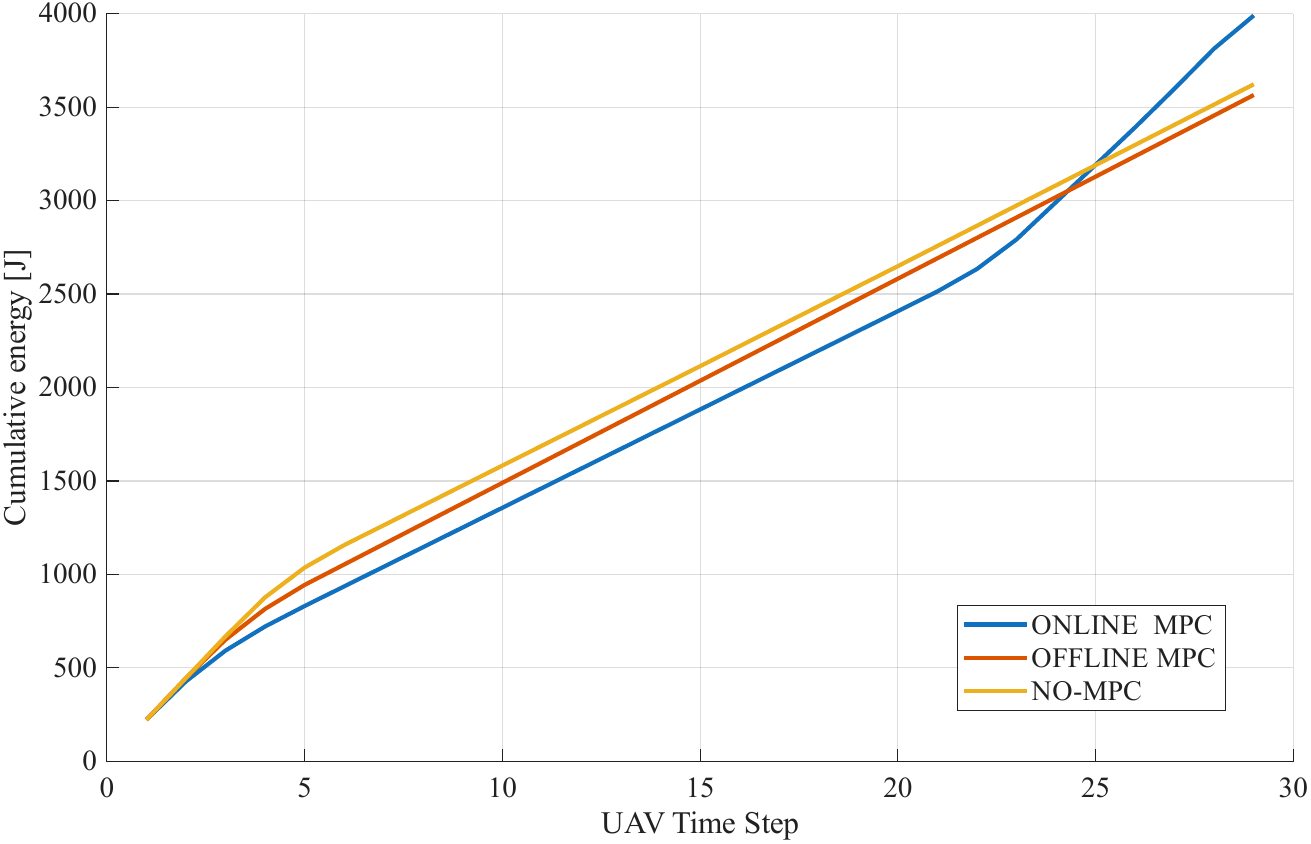}
    \captionof{figure}{Cumulative energy consumption under bounded disturbance of 6 m.}
    \label{fig:cumulative_energy_dev6}
\end{minipage}
\hfill
\begin{minipage}{0.45\textwidth}
    \centering
    \includegraphics[width=0.9\linewidth]{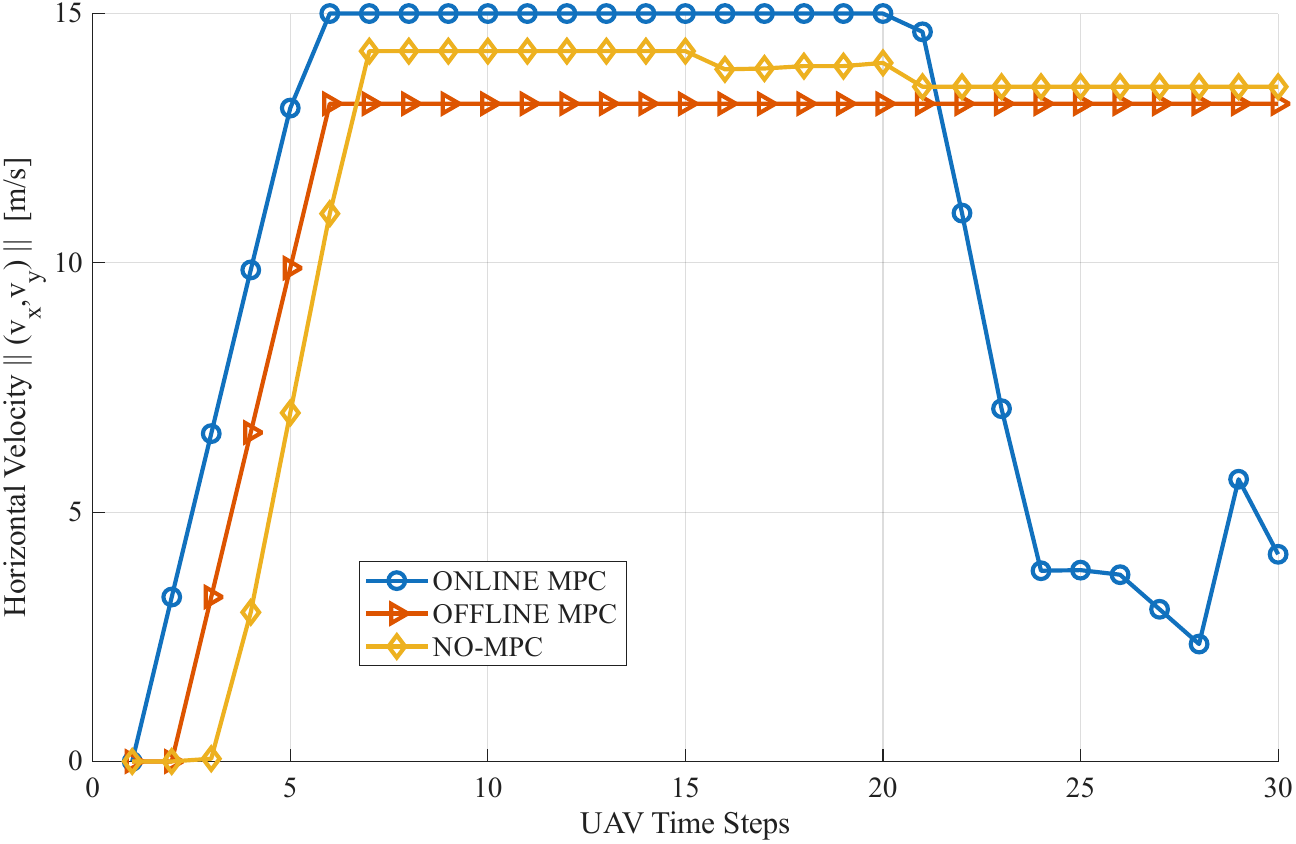}
    \captionof{figure}{Horizontal velocity evolution under different schemes.}
    \label{fig:horizontal_velocity_dev6}
\end{minipage}

\vspace{0.4cm}

% -------- Second Row --------
\begin{minipage}{0.45\textwidth}
    \centering
    \includegraphics[width=0.9\linewidth]{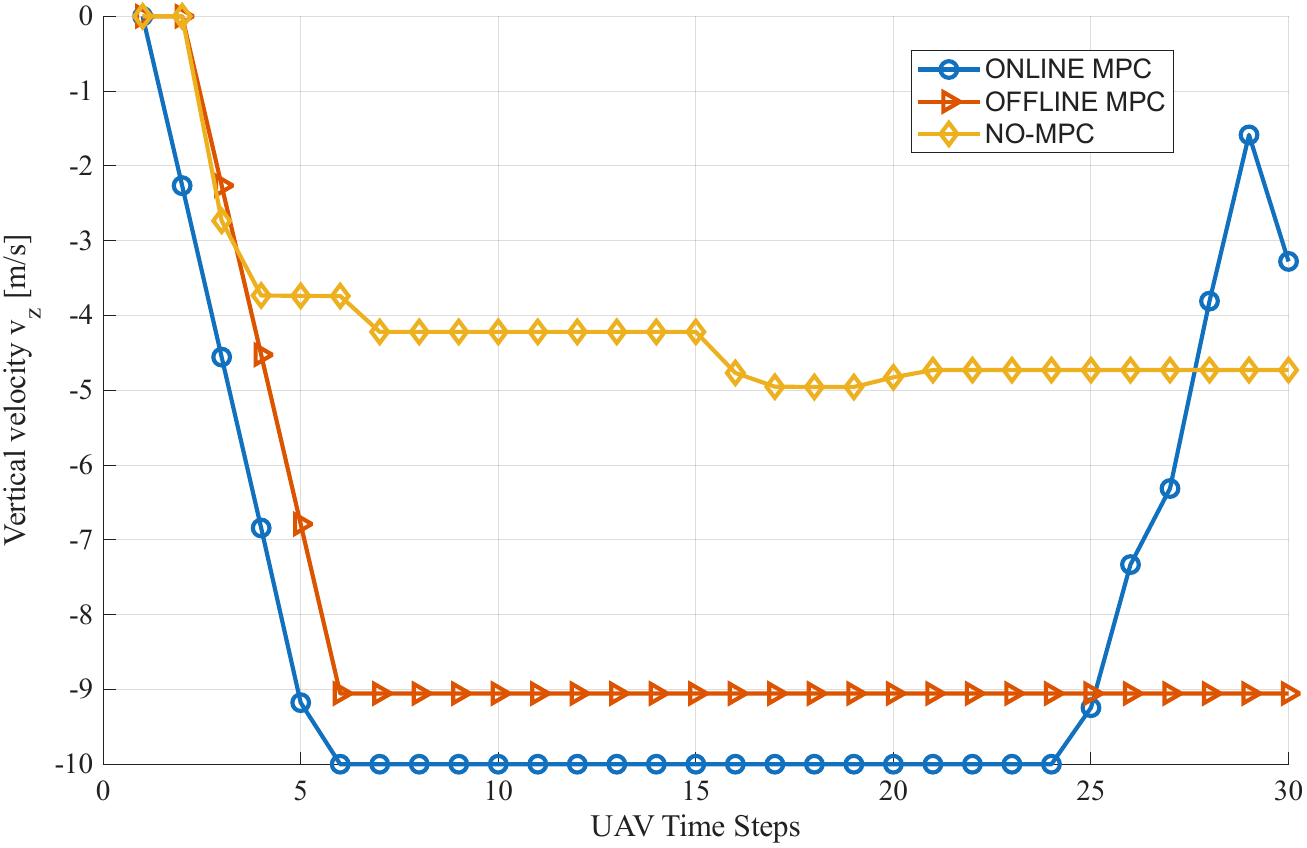}
    \captionof{figure}{Vertical velocity evolution of the UAV under different control schemes.}
    \label{fig:vertical_velocity_dev6}
\end{minipage}
\hfill
\begin{minipage}{0.45\textwidth}
    \centering
    \includegraphics[width=2.8 in]{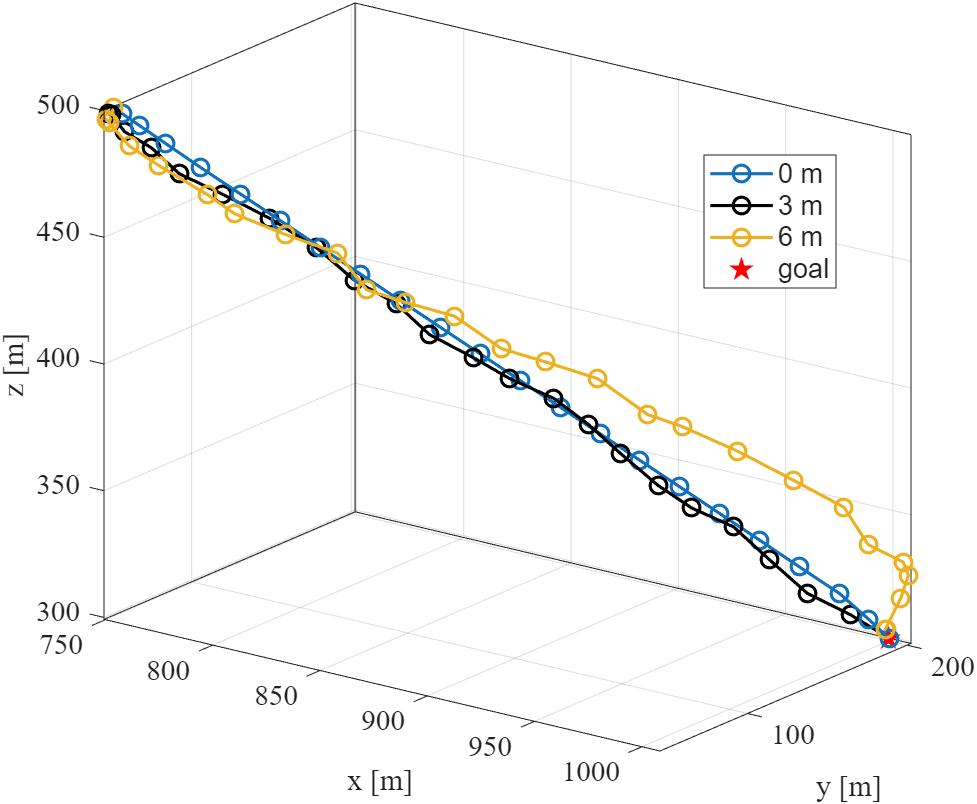}
    \captionof{figure}{UAV trajectories of the proposed online MPC under different bounded position disturbances (0 m, 3 m, and 6 m).}
    \label{fig:online_mpc_deviation}
\end{minipage}

\end{figure*}

Fig.~\ref{fig:cumulative_energy_dev6} presents the cumulative energy consumption under the same disturbance setting. Although the proposed online MPC introduces corrective adjustments during flight, its total energy expenditure remains comparable to the open-loop schemes. 
However, unlike the benchmarks, the online MPC successfully reaches the prescribed destination within the mission duration. These results demonstrate that the proposed closed-loop design achieves disturbance-resilient mission completion without incurring significant additional energy overhead.

Fig.~\ref{fig:horizontal_velocity_dev6} illustrates the evolution of the UAV horizontal velocity under the same disturbance setting. 
All schemes initially accelerate toward the destination, during which the horizontal speed approaches the maximum allowable value $V_{\max}$, indicating that the velocity constraint is active in the early stage to reduce travel time. While the open-loop schemes maintain nearly constant velocity profiles thereafter, the proposed online MPC adaptively adjusts the horizontal speed in later time steps to compensate for accumulated trajectory deviations. 
This adaptive motion regulation under constraint awareness enables accurate convergence within the prescribed mission duration.

Fig.~\ref{fig:vertical_velocity_dev6} depicts the vertical velocity evolution $v_z(t)$ under the same disturbance setting. From the initial and final UAV positions $r_A = [750, 50, 500]^T$\,m and $r_B = [1000, 200, 300]^T$\,m, the UAV is required to descend by $200$\,m during the mission. 
Accordingly, the vertical velocity remains predominantly negative, where negative $v_z(t)$ corresponds to downward motion under the upward-positive altitude convention. 
In the early stage, the descent rate approaches the maximum allowable magnitude $U_{\max}$, indicating active enforcement of the vertical speed constraint. 
While the open-loop schemes maintain nearly constant descent profiles, the proposed online MPC adaptively regulates $v_z(t)$ in later time steps to compensate for accumulated altitude deviations, thereby ensuring accurate convergence to the prescribed final altitude within the mission duration.

Fig.~\ref{fig:online_mpc_deviation} compares the UAV trajectories generated by the proposed online MPC under different bounded position disturbances of $0$\,m, $3$\,m, and $6$\,m. As the disturbance magnitude increases, the trajectory exhibits progressively larger deviations from the nominal path. Nevertheless, in all cases the UAV converges to the prescribed destination. This confirms that the receding-horizon re-optimization mechanism effectively compensates for increasing execution uncertainties and maintains trajectory robustness.

\begin{figure*}[!t]
\centering

% -------- First Row --------
\begin{minipage}{0.45\textwidth}
    \centering
    \includegraphics[width=0.9\linewidth]{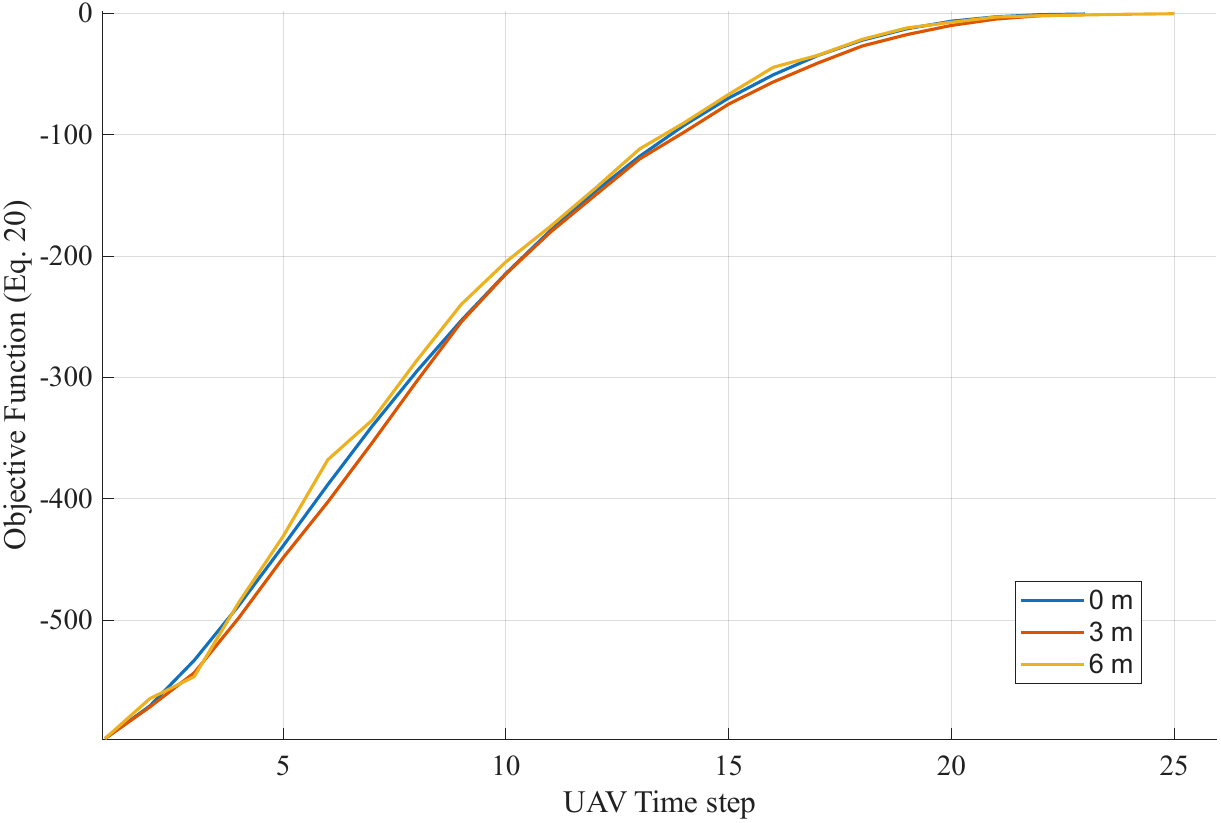}
    \captionof{figure}{Objective function evolution over UAV time steps.}
    \label{fig:objective_uav_timestep}
\end{minipage}
\hfill
\begin{minipage}{0.45\textwidth}
    \centering
    \includegraphics[width=0.9\linewidth]{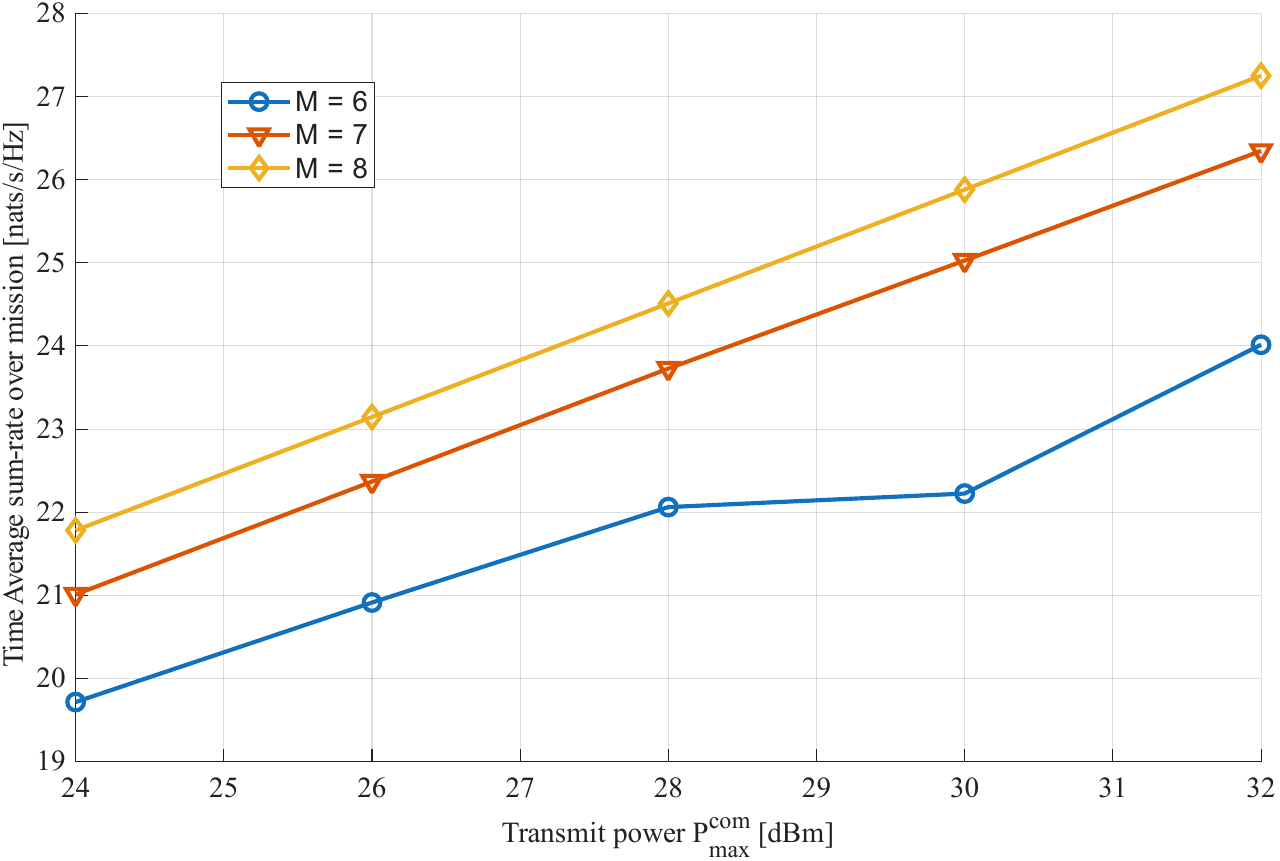}
    \captionof{figure}{Time-averaged URLLC sum-rate versus maximum transmit power $P^{com}_{{max}}$ for different antenna configurations.}
    \label{fig:qos_vs_power}
\end{minipage}

\vspace{0.4cm}

% -------- Second Row --------
\begin{minipage}{0.45\textwidth}
    \centering
    \includegraphics[width=0.9\linewidth]{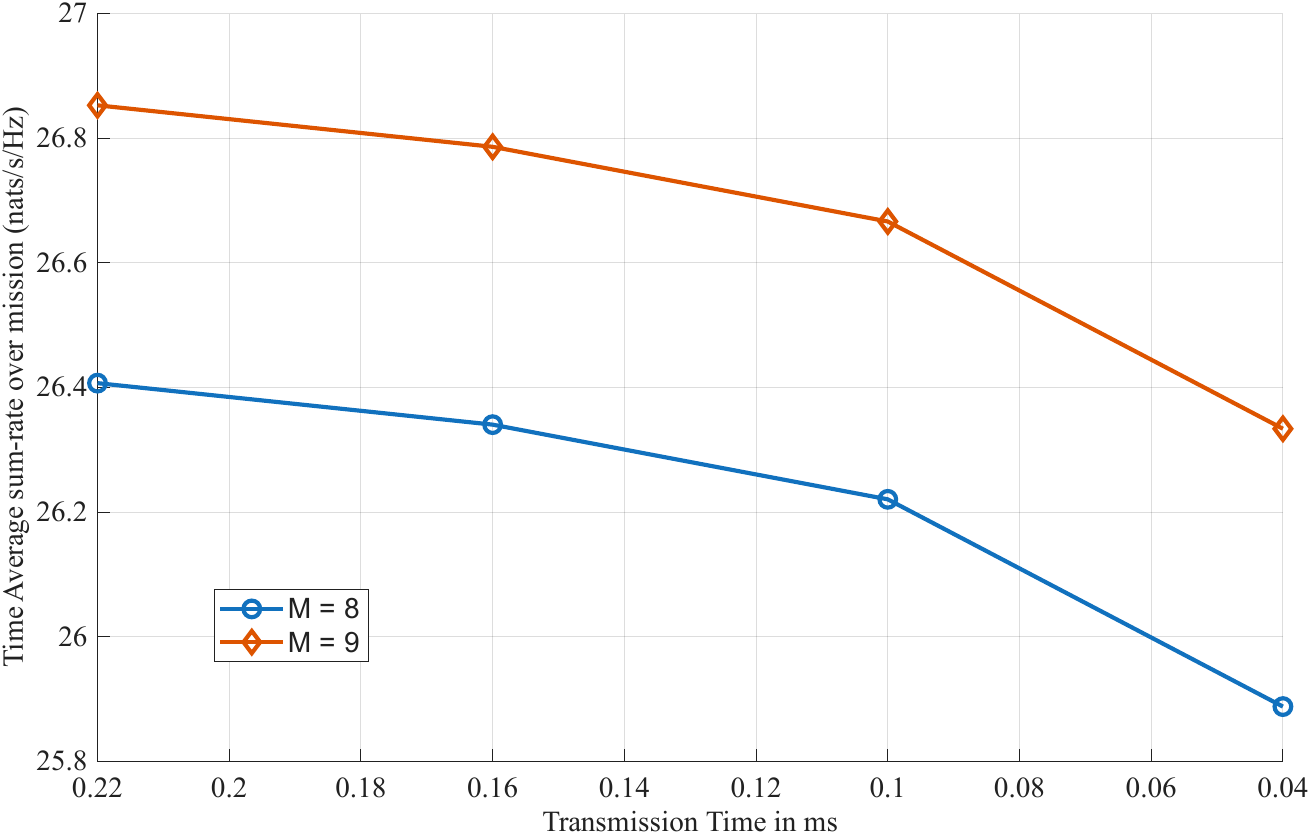}
    \captionof{figure}{Time-averaged sum-rate versus packet transmission time $T_t$ under FBL transmission.}
    \label{fig:sumrate_vs_time}
\end{minipage}
\hfill
\begin{minipage}{0.45\textwidth}
    \centering
    \includegraphics[width=0.9\linewidth]{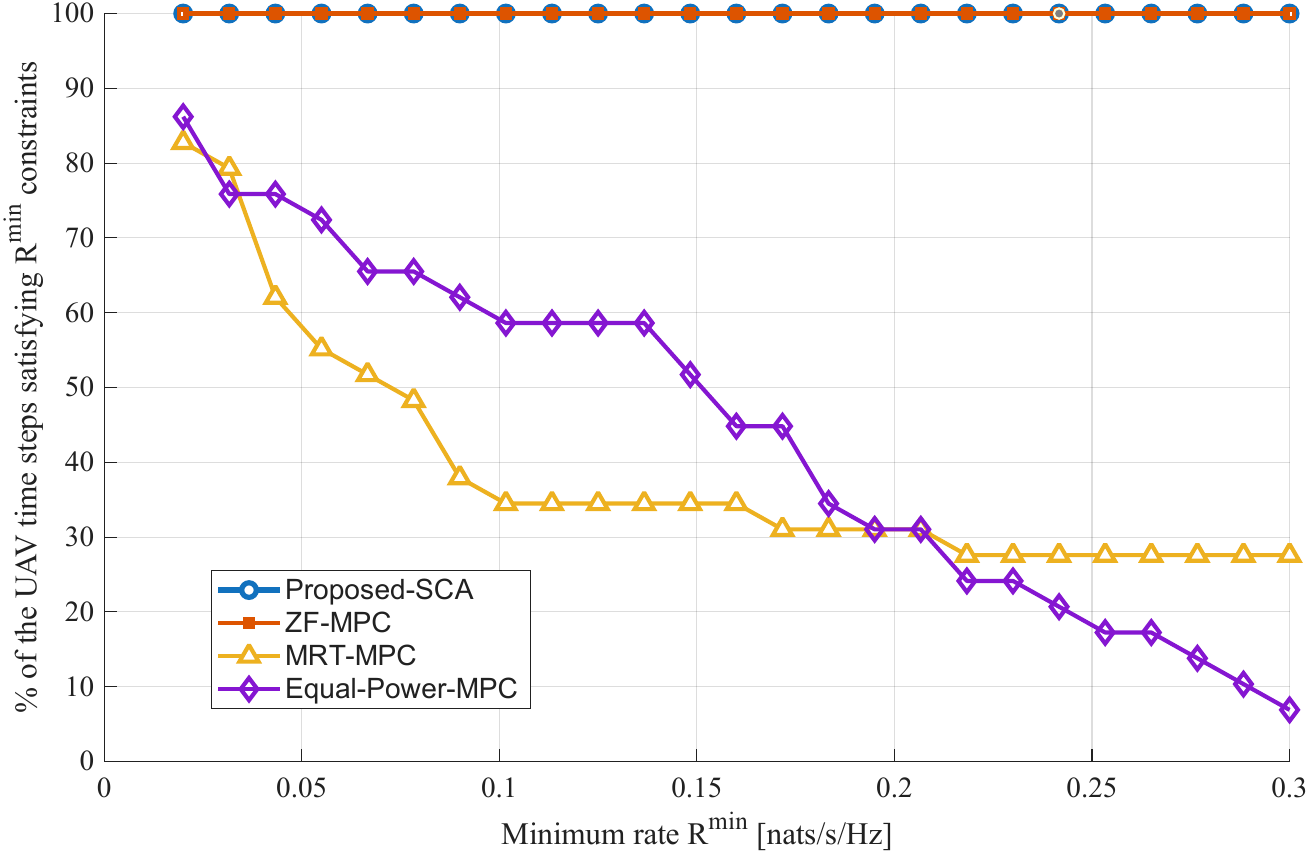}
    \captionof{figure}{Percentage of satisfied UAV time steps versus minimum required rate $R^{\min}$ under different beamforming strategies.}
    \label{fig:qos_rmin}
\end{minipage}

\end{figure*}

Fig.~\ref{fig:objective_uav_timestep} shows the evolution of the weighted objective function over UAV time steps under different trajectory disturbance magnitudes. The objective combines a sum-rate maximization term with penalties on the squared Euclidean distance to the destination and total power consumption. Because these components have different physical units and magnitudes, the weighting coefficients are chosen to normalize them to comparable numerical scales within the optimization. At the beginning of the mission, the UAV is far from the target and propulsion power is high; therefore, the weighted penalty terms dominate the scaled rate contribution, resulting in negative objective values. As the UAV approaches the destination, the squared distance penalty vanishes and propulsion consumption decreases significantly, leading to a monotonic increase of the objective. Near the final stage, the penalty terms become negligible and the objective is mainly determined by the normalized sum-rate term, which stabilizes close to zero due to the applied scaling rather than a reduction of the achievable sum-rate itself. The close alignment of the curves under different disturbance levels confirms the robustness and stability of the proposed MPC framework.

Fig.~\ref{fig:qos_vs_power} presents the time-averaged URLLC sum-rate versus the maximum transmit power for different antenna configurations. In this simulation, the number of users is set to $N=5$ to evaluate the communication performance under a denser multi-user scenario. The UAV trajectory is optimized for $M=6$ and kept fixed for $M=7$ and $M=8$, so performance differences isolate the pure communication gain from antenna scaling. As expected, the sum-rate increases monotonically with transmit power due to improved SINR under the FBL model. Moreover, increasing the number of antennas provides consistent performance gains across all power levels, owing to enhanced beamforming gain and interference suppression. These results demonstrate the significant impact of spatial degrees of freedom on URLLC performance, even without trajectory redesign.

Fig.~~\ref{fig:sumrate_vs_time} presents the time-averaged sum-rate versus transmission time under the FBL model. In this simulation, the number of users is set to $N=5$ to study the latency--spectral efficiency tradeoff under a denser multi-user scenario. Since the blocklength scales with transmission time, reducing $T$ shortens the effective coding length and increases the dispersion penalty, leading to performance degradation. This behavior reflects the fundamental latency–spectral efficiency tradeoff inherent to URLLC systems. Increasing the number of UAV antennas consistently improves performance across all transmission times due to enhanced beamforming gain and interference suppression, partially mitigating the rate loss caused by short-packet transmission.

Fig.~\ref{fig:qos_rmin} shows the percentage of UAV time steps satisfying the per time step URLLC constraint versus the minimum required rate $R_{\min}$. The UAV trajectory is obtained from the proposed MPC-based \textbf{P3} optimization and is kept fixed across all beamforming strategies; thus, performance differences arise purely from the transmission design. 
The proposed SCA-MPC and ZF-MPC maintain near-complete reliability over the tested $R_{\min}$ range, whereas MRT and equal-power beamforming degrade rapidly as the rate requirement increases. This gap is attributed to the ability of interference-aware designs to effectively suppress inter-user interference, which becomes particularly critical under FBL transmission where dispersion amplifies SINR sensitivity. These results demonstrate that interference-aware beamforming is essential for guaranteeing per time step URLLC feasibility.

\section{CONCLUSION}
This paper developed and validated a closed-loop predictive optimization framework for UAV-enabled URLLC systems operating under FBL transmission. Unlike conventional open-loop trajectory designs, the proposed online MPC architecture performs receding-horizon joint trajectory, velocity and beamforming optimization using real-time state feedback, thereby ensuring disturbance-resilient mission execution. Numerical results demonstrate that the closed-loop scheme guarantees convergence to the prescribed destination under the presence of UAV position disturbances, while offline and non-MPC benchmarks suffer trajectory drift and mission infeasibility. By explicitly enforcing per-time-step FBL rate constraints, the framework sustains URLLC reliability throughout the mission horizon. In particular, interference-aware beamforming (Proposed and ZF-based) achieves near-complete per time step feasibility even under stringent minimum rate requirements, whereas MRT and equal-power designs exhibit pronounced reliability degradation due to dispersion-sensitive SINR fluctuations, confirming that interference suppression is reliability-critical in the FBL regime. Moreover, the propulsion-aware convex reformulation enables robustness without significant energy overhead, maintaining mission-level energy consumption comparable to open-loop schemes. The results further quantify the impact of antenna scaling, transmit power, and transmission time, revealing the fundamental latency–spectral efficiency tradeoff and the role of spatial degrees of freedom in mitigating dispersion-induced performance loss. Overall, the proposed MPC-based framework provides a rigorous and practically implementable solution for reliability-centric UAV-enabled wireless networks in 5G and beyond systems. Future research directions include stochastic channel prediction within the MPC loop, distributed multi-UAV coordination, learning-assisted disturbance adaptation, and experimental validation on hardware-in-the-loop UAV communication platforms.
\appendices
\section{Derivation of the Concave Surrogate of the Shannon Term $C_n(\tau)$}
\label{appendix:shannon}
%------------------------------------------------
This appendix derives the concave lower bound used in
Eq.~(\ref{eq:Shannon_main}).
\subsection*{1.Taylor Expansion}
For the concave function $f(z)=\log(z)$, the first-order
Taylor expansion at $z^{(i)}>0$ provides the global lower bound
\begin{equation}
\log(z)
\le
\log(z^{(i)})
+
\frac{z-z^{(i)}}{z^{(i)}}.
\label{eq:A1}
\end{equation}

Applying Eq.~\eqref{eq:A1} to the two logarithmic terms of
$C_n(\tau)=\log(S_n+I_n)-\log(I_n)$ at
$(S_n^{(i)}, I_n^{(i)})$
and summing the resulting inequalities yields
\begin{equation}
C_n(\tau)
\ge
C_n^{(i)}(\tau)
+
\frac{S_n+I_n-S_n^{(i)}-I_n^{(i)}}
{S_n^{(i)}+I_n^{(i)}}
-
\frac{I_n-I_n^{(i)}}{I_n^{(i)}},
\label{eq:A2}
\end{equation}
Define the increment term
\begin{equation}
L_n^{(i)}(\tau)
\triangleq
\frac{S_n + I_n - S_n^{(i)} - I_n^{(i)}}
     {S_n^{(i)} + I_n^{(i)}}
-
\frac{I_n - I_n^{(i)}}{I_n^{(i)}} .
\end{equation}

After algebraic manipulation, this can be written as
\begin{equation}
L_n^{(i)}(\tau)
=
\frac{S_n(\tau)}{I_n^{(i)}(\tau)}
-
\frac{S_n^{(i)}(\tau)}
     {I_n^{(i)}(\tau)\big(S_n^{(i)}(\tau)+I_n^{(i)}(\tau)\big)}
\big(S_n(\tau)+I_n(\tau)\big).
\label{eq:Ln_i}
\end{equation}

\subsection*{2. Substitution of the Linearized Signal Term}

Since $S_n(\tau)=|\mathbf h_n^H(\tau)\mathbf w_n(\tau)|^2$
is convex in $\mathbf w_n(\tau)$,
its first-order Taylor expansion provides an affine lower bound given in Eq.~(\ref{eq:signal_lb_TVT}). Substituting this affine bound into Eq.~\eqref{eq:Ln_i}
yields the explicit affine expression of $L_n^{(i)}(\tau)$
reported in Eq.~\eqref{eq:L_main_TVT}.

\subsection*{3. Trust-Region Condition}

To ensure validity of the linearize signal term, the following trust region condition is enforced: 
\begin{equation}
2\Re\!\left\{
\big(\mathbf h_n^H(\tau)\mathbf w_n^{(i)}(\tau)\big)^*
\mathbf h_n^H(\tau)\mathbf w_n(\tau)
\right\}
-
\big|\mathbf h_n^H(\tau)\mathbf w_n^{(i)}(\tau)\big|^2
>0.
\label{eq:A3}
\end{equation}
%------------------------------------------------
%-------------------------------------------------
\section{Derivation of the Convex Surrogate of the FBL Penalty Term ${D_n(\tau)}$}
\label{appendix:fbl}

This appendix derives the affine upper bound of the FBL penalty term used in Section~III-A-b.

%-------------------------------------------------
\subsection*{1. First-Order Upper Bound of $\sqrt{V_n(\tau)}$}

Since the function $\sqrt{x}$ is concave for $x>0$, its first-order Taylor expansion at $V_n^{(i)}(\tau)$ gives the global upper bound
\begin{equation}
\sqrt{V_n(\tau)}
\le
\sqrt{V_n^{(i)}(\tau)}
+
\frac{1}{2\sqrt{V_n^{(i)}(\tau)}}
\left(
V_n(\tau)-V_n^{(i)}(\tau)
\right).
\end{equation}

Rearranging the above expression yields
\begin{equation}
\sqrt{V_n(\tau)}
\le
\frac{V_n(\tau)}{2\sqrt{V_n^{(i)}(\tau)}}
+
\frac{1}{2}\sqrt{V_n^{(i)}(\tau)}.
\label{eq:sqrt_x}
\end{equation}
Substituting the expression of $V_n(\tau)$ from Eq.~(\ref{eq:V_main}) in Eq.~(\ref{eq:sqrt_x}) gives
\begin{equation}
\sqrt{V_n(\tau)}
\le
B_n^{(i)}
-
A_n^{(i)}
\frac{I_n(\tau)^2}{\left(S_n(\tau)+I_n(\tau)\right)^2}.
\label{sqrt_Vn}
\end{equation}
where,
\begin{equation}
A_n^{(i)}=\frac{1}{2\sqrt{V_n^{(i)}(\tau)}}, \qquad
B_n^{(i)}=A_n^{(i)}+\frac{1}{2}\sqrt{V_n^{(i)}(\tau)} .
\end{equation}
%-------------------------------------------------
\subsection*{2. Decomposition of the Fractional Term}

Consider the fractional term
\begin{equation}
\frac{I_n(\tau)^2}{(S_n(\tau)+I_n(\tau))^2}.
\end{equation}

We decompose it as
\begin{equation}
\frac{I_n^2}{(S_n+I_n)^2}
=
\frac{I_n^2}{S_n+I_n}
\cdot
\frac{1}{S_n+I_n}.
\end{equation}

For the second factor, applying first-order Taylor expansion of
$f(x)=1/x$ at $x^{(i)}=S_n^{(i)}+I_n^{(i)}$ yields
\begin{equation}
\frac{1}{S_n+I_n}
\ge
\frac{2}{S_n^{(i)}+I_n^{(i)}}
-
\frac{S_n+I_n}{(S_n^{(i)}+I_n^{(i)})^2},
\end{equation}
under the trust-region condition
\begin{equation}
S_n+I_n \le 2(S_n^{(i)}+I_n^{(i)}).
\end{equation}

For the first factor, we now employ the inequality in~\cite[Eq.~(21)]{7845589}, which gives the concave lower bound
\begin{equation}
\frac{I_n^2}{S_n+I_n}
\ge
\frac{2 I_n^{(i)}}{S_n^{(i)}+I_n^{(i)}} I_n
-
\frac{(I_n^{(i)})^2}{(S_n^{(i)}+I_n^{(i)})^2}(S_n+I_n).
\label{fractional_term}
\end{equation}
under the trust-region condition
\begin{equation}
\frac{(S_n+I_n)}{(S_n^{(i)}+I_n^{(i)})} \le\frac{2I_n}{I_n^{(i)}}
\end{equation}

Using the above approximations for both factors and performing algebraic simplification, the fractional term admits the following affine upper bound
\begin{equation}
\frac{I_n^2}{(S_n+I_n)^2}
\le
\alpha_n^{(i)} I_n
+
\beta_n^{(i)} (S_n+I_n)
+
\psi_n^{(i)},
\label{Final_fraction}
\end{equation}
where the coefficients
$\alpha_n^{(i)}$,
$\beta_n^{(i)}$,
and
$\psi_n^{(i)}$
are defined in Section~III-A-b.

%-------------------------------------------------
\subsection*{3. Substitution into the Dispersion Bound}

Substituting the affine upper bound of the fractional term
in  Eq.~\eqref{Final_fraction}
into the first-order expansion of $\sqrt{V_n(\tau)}$
in Eq.~\eqref{sqrt_Vn}, and multiplying by $c_n$,
yields the affine surrogate
$D_n(\tau)\le \widetilde{D}_n^{(i)}(\tau)$
reported in Section~III-A-b.
%-------------------------------------------------

\section{Convex Reformulation of the Propulsion Power Model}

The propulsion power model in Eq. (6) contains the induced-power component
\begin{equation}
\phi(\mathbf v_h)
=
\frac{W^2}{\sqrt{2}\rho S}
\Big(
\|\mathbf v_h\|_2^2
+
\sqrt{\|\mathbf v_h\|_2^4+4V_h^4}
\Big)^{-1/2},
\label{eq:C1}
\end{equation}
where $\mathbf v_h=[v_x,\,v_y]^T$ denotes the horizontal velocity.

Define
\begin{equation}
\Psi(\mathbf v_h)
=
\|\mathbf v_h\|_2^2
+
\sqrt{\|\mathbf v_h\|_2^4+4V_h^4}.
\label{eq:C2}
\end{equation}

Let $s=\|\mathbf v_h\|_2^2 \ge 0$. Then
\begin{equation}
\frac{d}{ds}
\left(
s+\sqrt{s^2+4V_h^4}
\right)
=
1+\frac{s}{\sqrt{s^2+4V_h^4}},
\label{eq:C3}
\end{equation}
\begin{equation}
\frac{d^2}{ds^2}
\left(
s+\sqrt{s^2+4V_h^4}
\right)
=
\frac{4V_h^4}{(s^2+4V_h^4)^{3/2}},
\label{eq:C4}
\end{equation}
which are positive for all $s\ge0$. Hence, $\Psi(\mathbf v_h)$ is convex in $\mathbf v_h$.

At SCA iteration $i$, the first-order lower bound of $\Psi(\mathbf v_h)$ at $\mathbf v_h^{(i)}$ is
\begin{equation}
\Psi(\mathbf v_h)
\ge
a_i
+
\mathbf g_i^\top
\big(
\mathbf v_h-\mathbf v_h^{(i)}
\big),
\label{eq:C5}
\end{equation}
where
\begin{equation}
a_i
=
\Psi(\mathbf v_h^{(i)}),
\label{eq:C6}
\end{equation}
\begin{equation}
\mathbf g_i
=
2
\left(
1+
\frac{\|\mathbf v_h^{(i)}\|_2^2}
{\sqrt{\|\mathbf v_h^{(i)}\|_2^4+4V_h^4}}
\right)
\mathbf v_h^{(i)}.
\label{eq:C7}
\end{equation}

Since $x^{-1/2}$ is convex and monotonically decreasing for $x>0$, composing it with \eqref{eq:C5} yields the convex upper bound
\begin{equation}
\phi(\mathbf v_h)
\le
\frac{W^2}{\sqrt{2}\rho S}
\Big(
a_i
+
\mathbf g_i^\top
\big(
\mathbf v_h-\mathbf v_h^{(i)}
\big)
\Big)^{-1/2}.
\label{eq:C8}
\end{equation}

The resulting convex surrogate propulsion power is
\begin{equation}
\begin{aligned}
P_{\mathrm{prop}}^{\mathrm{ub}}(\mathbf v)
&=
\frac{W^2}{\sqrt{2}\rho S}
\Big(
a_i
+
\mathbf g_i^\top
\big(
\mathbf v_h-\mathbf v_h^{(i)}
\big)
\Big)^{-1/2} \\
&\quad
+ W \max\big(v_z(t), 0\big)
+ \frac{\zeta\rho S}{8}\|\mathbf v_h\|_2^3 .
\end{aligned}
\label{eq:C9}
\end{equation}

Accordingly, the total power constraint C6 in the trajectory subproblem \textbf{(P3)} is conservatively reformulated as
\begin{equation}
P_{\mathrm{prop}}^{\mathrm{ub}}(\mathbf v(\tau))
+
P_{\mathrm{com}}(\tau)
\le
P_{\max},
\quad
\forall \tau\in\mathcal T_t,
\label{eq:C10}
\end{equation}

The surrogate constraint \eqref{eq:C10} is convex in $\mathbf v(\tau)$ and tight at the linearization point, thereby ensuring tractability of the propulsion power constraint within the SCA-based MPC framework.
%--------------------------------------------

\bibliographystyle{IEEEtran}
\bibliography{refs.bib}

%\printbibliography

\end{document}